\newif\ifjmr

\newif\ifreview

\newcommand{\papertitle}{Your MMM is Broken: \\ Identification of Nonlinear and Time-varying Effects in Marketing Mix Models}

\ifjmr
    \documentclass[12pt]{article}
    \input{jmr_style}
    \title{\papertitle}
\else
    \documentclass[11pt]{article}
\usepackage[pdftex,letterpaper,left=1in,top=1in,bottom=1in,right=1in]{geometry}
\usepackage{afterpage, setspace} 
\usepackage[titletoc,title]{appendix}
\usepackage[dotinlabels]{titletoc}
\usepackage{titlesec, marginnote, ftnxtra, todonotes}
\usepackage[round,longnamesfirst]{natbib}
\setcitestyle{aysep={}} 
\usepackage[nottoc]{tocbibind} 
\usepackage{multibib}
\newcites{latex}{References (Appendix)} 
\AtBeginEnvironment{appendices}{\crefalias{section}{appendix}\crefalias{subsection}{appendix}\crefalias{subsubsection}{appendix}}

\usepackage{amsmath, amssymb, mathtools, amsfonts, amsthm, verbatim, nicefrac, bm, bbm, fancyhdr, array, listings, xcolor, enumitem, thmtools, subcaption, pdflscape, pgffor, csquotes, listings, lstbayes, longtable, soul}
\usepackage{comment}
\usepackage{booktabs, tablefootnote, makecell, multirow}
\usepackage[referable]{threeparttablex}
\setlist{noitemsep,parsep=6pt,partopsep=0pt,topsep=0pt}
\usepackage{graphicx, color, xcolor}
\usepackage{pdflscape}
\usepackage{adjustbox}
\usepackage{ragged2e}
\usetikzlibrary{bayesnet}
\usepackage{authblk}

\definecolor{main}{rgb}{0.05,0.25,0.56}
\definecolor{dark-red}{rgb}{0.4,0.15,0.15}
\definecolor{dark-blue}{rgb}{0.05,0.25,0.56}
\definecolor{light-blue}{rgb}{.2,1,1}
\definecolor{medium-blue}{rgb}{0,0,0.5}
\definecolor{dark-green}{rgb}{0,.5,0}
\usepackage[colorlinks, pagebackref=false]{hyperref}
\usepackage{caption}
\usepackage[nameinlink]{cleveref}
\hypersetup{
    citecolor={dark-blue},
    bookmarksnumbered=true,
    urlcolor=dark-red,
    linkcolor=dark-blue
}

\creflabelformat{equation}{#2#1#3}
\usepackage{mathpazo}
\usepackage[margin=15pt,font=small,labelfont={bf,sf}]{caption}
\usepackage{helvet}
\usepackage{sectsty}
\allsectionsfont{\sffamily\color{dark-blue}}
\usepackage{colortbl}

\newcolumntype{L}[1]{>{\hangindent=1ex \raggedright\let\newline\\\arraybackslash\hspace{0pt}}p{#1}}
\newcolumntype{C}[1]{>{\centering\let\newline\\\arraybackslash\hspace{0pt}}m{#1}}
\newcolumntype{R}[1]{>{\raggedleft\let\newline\\\arraybackslash\hspace{0pt}}m{#1}}
\newcolumntype{H}{>{\setbox0=\hbox\bgroup}c<{\egroup}@{}}

{

}

\newcommand{\reals}{\mathbb{R}}

\newcommand{\x}{{\mathbf{x}}}
\newcommand{\xbar}{{\bar{x}}}

\newcommand{\N}{{\mathcal{N}}}

\newcommand{\btheta}{\boldsymbol{\theta}}

\newcommand{\tcaption}[2]{\caption{\textbf{#1} \\ #2}}
    \title{\textbf{\sffamily\color{dark-blue} \papertitle}\thanks{Ryan Dew (\texttt{ryandew@wharton.upenn.edu}) is an Assistant Professor of Marketing and the Govil Family Faculty Scholar at the Wharton School, University of Pennsylvania. Nicolas Padilla (\texttt{npadilla@london.edu}) is an Assistant Professor of Marketing at London Business School. Anya Shchetkina (\texttt{annashch@wharton.upenn.edu}) is a doctoral student at the Wharton School, University of Pennsylvania. Authors contributed equally and are listed alphabetically. The authors thank Christophe Van den Bulte and Eric Bradlow for helpful feedback, as well as seminar participants at Emory University, ESADE, and HKUST. Soham Mahadik provided excellent research assistance.}}
\fi

\input{xr}
\ifreview
    \author{Authors masked for review.}
\else
    \author[1]{Ryan Dew}
    \author[2]{Nicolas Padilla}
    \author[1]{Anya Shchetkina}
    \affil[1]{The Wharton School, University of Pennsylvania}
    \affil[2]{London Business School}
\fi

\begin{document}
    
\maketitle

\begingroup
\justifying
\begin{abstract}
\noindent Recent years have seen a resurgence in interest in marketing mix models (MMMs), which are aggregate-level models of marketing effectiveness. Often these models incorporate nonlinear effects, and either implicitly or explicitly assume that marketing effectiveness varies over time. In this paper, we show that such effects are often not separately identifiable: while certain data patterns may be suggestive of nonlinear effects, such patterns may also emerge under simpler models with time-varying effects. Problematically, nonlinearities and dynamics suggest fundamentally different optimal marketing allocations. We examine this identification issue through theory and simulations, describing the conditions under which conflation between the two types of models is likely to occur. We show that conflating the two types of effects is especially likely in the presence of autocorrelated marketing variables, which are common in practice, especially given the widespread use of stock variables to capture long-run effects of advertising. We illustrate these ideas through numerous empirical applications to real-world marketing mix data, showing the prevalence of the conflation issue in practice. Finally, we show how marketers can avoid this conflation, by designing experiments that strategically manipulate spending in ways that pin down model form. 
\end{abstract}
\endgroup


\newpage
\ifjmr
\setstretch{1.65}
\else
\onehalfspacing
\fi

\ifjmr
\else
\section{Introduction} 
\fi

Recent years have seen a seismic shift in the marketing measurement landscape: whereas advertisers previously relied on third-party cookies to track consumers online, and thus make attributions about how their ads are driving the path-to-purchase \citep[e.g.,][]{li2014attributing, li2016attribution, zantedeschi2017measuring, berman2018beyond}, new privacy laws and restrictions have made doing so nearly impossible \citep{cui2021informational}. This, combined with a growing understanding that such individual-level attribution models are often fundamentally flawed \citep[e.g.,][]{gordon2023close,tian2024cookie}, has shifted attention back to a classic marketing measurement tool: aggregate-level marketing mix models, or MMMs \citep{hanssens2003market}.\footnote{Often in practice marketing mix models exclusively use advertising expenditures across various channels as the independent variables, ignoring other aspects of the marketing mix. In these cases, these types of models are referred to as \textit{media mix models}.} This shift in industry focus is reflected in both a flurry of research being done in practice, including from companies like Google and Uber \citep[e.g.,][]{jin2017bayesian, wang2017hierarchichal, ng2021bayesian}, and in the open source tools being released by these companies, like Google's Meridian \citep{meridian} and Meta's Robyn \citep{robyn}, all of which implement highly sophisticated MMMs.

While the motivating force behind this refocusing is new, MMMs are not: the concept of the marketing mix dates back to \cite{borden1964concept}, and the idea of modeling the effects of the marketing mix through aggregate measures is also quite old \citep{quandt1964estimating, palda1965measurement}. In the ensuing decades, many topics in this space have been studied, including long-term effects of advertising, especially as measured through lag and stock variables \citep[e.g.,][]{mann1975optimal}, time-dependent effects, especially using models with time-varying coefficients \citep[e.g.,][]{parsons1975product,winer1979analysis}, nonlinear effects, including diminishing returns \citep[e.g.,][]{vidale1957operations}, how to optimize marketing spend in the face of such forces \citep[e.g.,][]{sasieni1971optimal}, and, more recently, issues of endogeneity in the specification of such models \citep[e.g.,][]{luan2010forecasting}. What is new is the sophistication and scale of the MMMs being developed and enabled through these open-source libraries, which implement many of these classic ideas in the context of complex Bayesian hierarchical models, leveraging modern probabilistic programming languages for estimation \citep{wang2017hierarchichal}.

\begin{figure}
    \centering
    \includegraphics[width=0.5\textwidth]{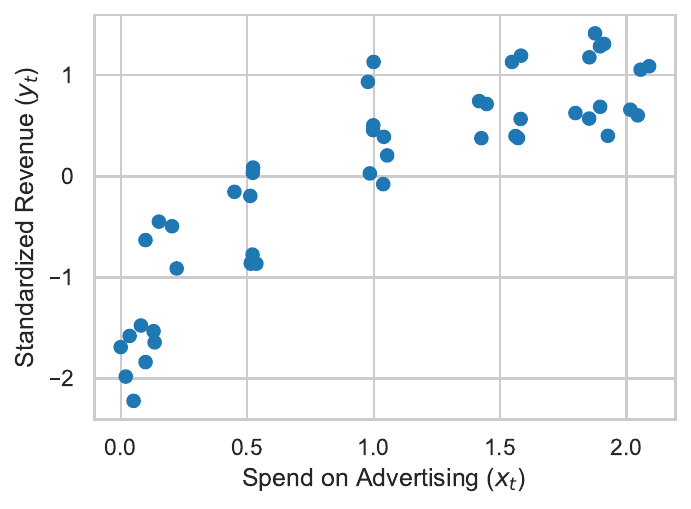}
    \tcaption{Illustrative Example}{A simulated example showing a seemingly nonlinear effect of advertising on revenue.}
    \label{fig:intro_example_data}
\end{figure}

\begin{figure}
    \centering
    \includegraphics[width=\textwidth]{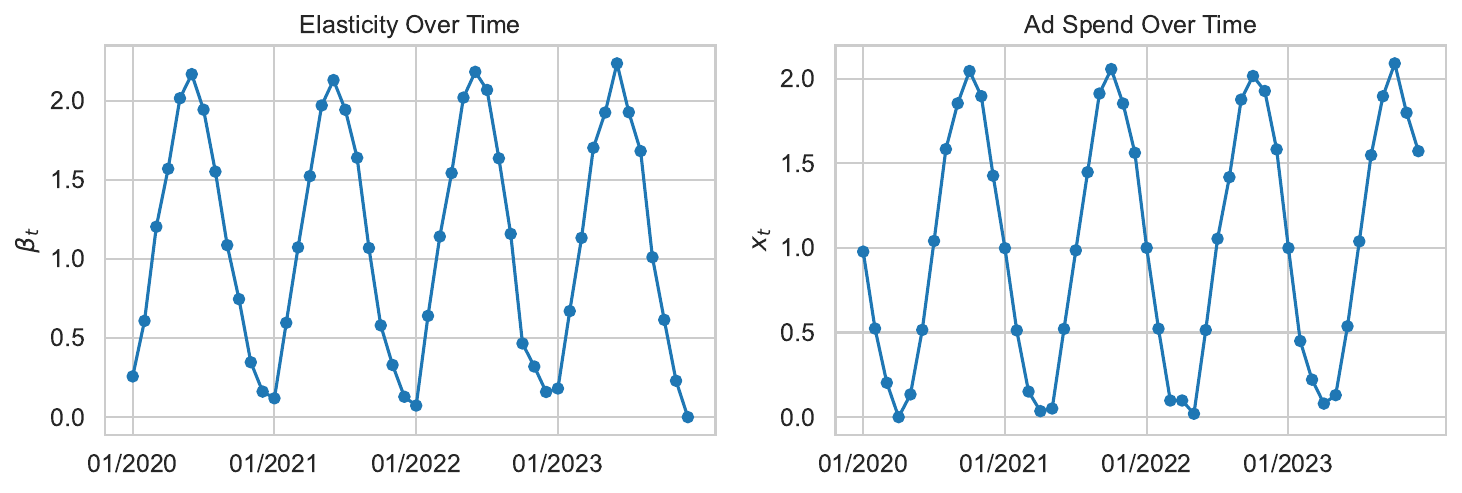}
    \tcaption{True Data Generating Process}{The true data generating process for \Cref{fig:intro_example_data}, which is a linear time-varying effects model, $y_t = \beta_t x_t + \varepsilon_t$, with $\beta_t$ evolving cyclically as shown in the left panel, and $x_t$ set following $\beta_t$, as shown in the right panel.}
    \label{fig:intro_example_dgp}
\end{figure}

Despite the sophistication of modern MMMs, the type of data used has remained fundamentally unchanged: time series of aggregate measures of spending on various marketing channels, and aggregate response measures like revenue or customers acquired. This relatively simple type of data, paired with increasingly complex models, raises a fundamental question as to which types of effects posited by practitioners can actually be reliably estimated. To illustrate, consider the (synthetic) example plotted in \Cref{fig:intro_example_data}, of revenue versus spend on advertising. One cornerstone of both modern and classic MMMs is the estimation of nonlinear effects, aimed at capturing potential inflection points and diminishing returns in the effects of advertising. In \Cref{fig:intro_example_data}, such effects seem especially plausible: indeed, a flexible nonlinear model, of the form $y_t = f(x_t) + \varepsilon_t$, is able to capture most of the variation in the data. In fact, however, the true data-generating process of this example is a \textit{linear} model, where the effect of advertising varies smoothly over time: $y_t = \beta_t x_t + \varepsilon_t$. Notably, to generate this example, we assumed that the firm sets spending with knowledge of how $\beta_t$ is evolving, such that $x_t$ noisily mirrors $\beta_t$ with some lag, as shown in \Cref{fig:intro_example_dgp}. While somewhat contrived, this time-varying data-generating process mirrors other common MMM practices. For instance, it is common for firms to estimate their MMMs on rolling windows, doing periodic ``refreshes'' to understand how the parameters of the MMM have changed, and adjusting spending accordingly, thereby implicitly assuming time-varying coefficients. Other MMMs explicitly allow for time-varying coefficients, both in linear and nonlinear (e.g., log-log) models. That both the nonlinear and time-varying models fit the data equally well is a problem, as the two sets of assumptions have fundamentally different implications for optimal spending: under the nonlinear model, the firm should set spending at a constant, optimal level, dictated by the response curve, while in the time-varying model, the firm's spending should depend on the evolution of $\beta_t$. 

In this paper, we investigate when nonlinear and time-varying effects can be separately identified from standard marketing mix data. Through theory, extensive simulations, and real-world MMM applications, we show that, frequently, they cannot be. In particular, given sufficient flexibility in their specification, these two classes of models can often approximate one another, implying that standard metrics of model fit, including cross-validation, cannot indicate which is correct. This problem exists even when the data used in constructing the models are exogenous (i.e., when $x_t$ is independent of $\varepsilon_t$). We further show that this conflation is more likely to occur when spending is set in an autoregressive way, which is not only common in practice (e.g., when decision-makers set current spending as a function of the past), but also commonly \textit{induced} by modeling the effect of $x_t$ through a stock variable, like AdStock, which is a staple in both classic and modern MMMs \citep[e.g.,][]{hanssens2003market,zantedeschi2017measuring,meridian,robyn}. As hinted by the illustration above, such conflation is not merely a statistical curiosity: we show that nonlinear and time-varying models can have fundamentally different implications for setting optimal spending. This is because, while the two models give equivalent predictions under status quo spending patterns, under \textit{intervention}, their predictions may differ. This difference also suggests a solution to the conflation problem: specifically, we show that marketers can implement tests to identify the two types of effects, by means of carefully designed experiments. These experiments are akin to a specific type of ``incrementality test,'' which are already common in industry for measuring the effect of spending on a marketing channel. 

A crucial aspect of our analysis is being able to both simulate from and flexibly estimate nonlinear and time-varying effects. To do so, we develop a Bayesian nonparametric framework based on Gaussian processes. Gaussian processes, or GPs, provide a means of simulating functions with differing properties, including differing levels of smoothness, which is important for highlighting some of our key theoretical results via simulation. As the basis of a Bayesian marketing mix model, GPs can also be used to estimate flexible nonlinear effects. This flexibility enables us to capture both the standard types of nonlinearities often used in practice, and more general ones. Moreover, when applied to the coefficients of a regression model, GPs can also capture their smooth variation over time.\footnote{Theoretically, this framework can also simultaneously model \textit{both} time-varying \textit{and} nonlinear effects, though we do not explore such a specification in-depth here.}

The rest of the paper is organized as follows: first, we review the underpinnings of marketing mix models, focusing on the specifications used in contemporary marketing practice. Then, we introduce at the most general level the nonlinear and dynamic models of interest, and show analytically how (and when) they can approximate one another. We then describe our GP-based framework for simulating and estimating these effects. We use this framework to conduct extensive simulations, which illustrate the potential extent of conflation, as well as its moderators and implications. Then, we analyze a series of real-world marketing mix datasets, including the classic Lydia Pinkham data \citep{palda1965measurement}, the dietary weight control product sales dataset analyzed in \cite{bass1972testing}, and four more recent examples constructed from the NielsenIQ Retail Scanner and Nielsen AdIntel data,\footnote{Disclaimer: Researcher(s)' own analyses calculated (or derived) based in part on (i) retail measurement/consumer data from Nielsen Consumer LLC ("NielsenIQ"); (ii) media data from The Nielsen Company (US), LLC ("Nielsen"); and (iii) marketing databases provided through the respective NielsenIQ and the Nielsen Datasets at the Kilts Center for Marketing Data Center at The University of Chicago Booth School of Business.

The conclusions drawn from the NielsenIQ and Nielsen data are those of the researcher(s) and do not reflect the views of Nielsen. Nielsen is not responsible for, had no role in, and was not involved in analyzing and preparing the results reported herein.} highlighting the prevalence of conflation in practice. Finally, we introduce a series of tests, based on intervening on spending, that can be used to diagnose which effects are most relevant in practice, and thus separately identify them as part of an MMM. We conclude with a summary and directions for future research.

\section{Background}

The literature on aggregate-level marketing mix models is vast, and we do not provide an exhaustive review here. Interested readers can find an extensive review of the history and foundation of these models in \cite{hanssens2003market}. There is also a long literature in marketing that looks more narrowly at how advertising impacts sales \citep[e.g.,][]{dekimpe2007advertising, shapiro2021tv}, which is often the focal use-case of MMMs in practice, and which we will also not exhaustively review. Instead, we focus on providing background on the key effects of interest in our analysis: nonlinear effects, carryover effects, and time-varying effects. 

\subsection{Nonlinear Effects}

The response of revenue to marketing actions, particularly advertising, has long been hypothesized to be nonlinear \citep[e.g.,][]{vidale1957operations, rao1975advertising, simon1980shape, vakratsas2004shape}. That is, as described previously,
\begin{equation}
    y_t = f(x_t) + \varepsilon_t, \label{eq:uni_nl_model}
\end{equation}
where for simplicity, we focus on spend in only a single marketing channel (e.g., advertising), $x_t$, and $y_t$ is an outcome of interest, typically profit or revenue. The most commonly posited functional relationship between advertising and revenue is an S-shaped function, which accounts for two effects: a threshold effect, wherein advertising is only effective if spend exceeds a certain threshold; and diminishing returns \citep[e.g.,][]{cannon2002beyond}. 

\begin{figure}
    \centering
    \includegraphics[width=0.5\textwidth]{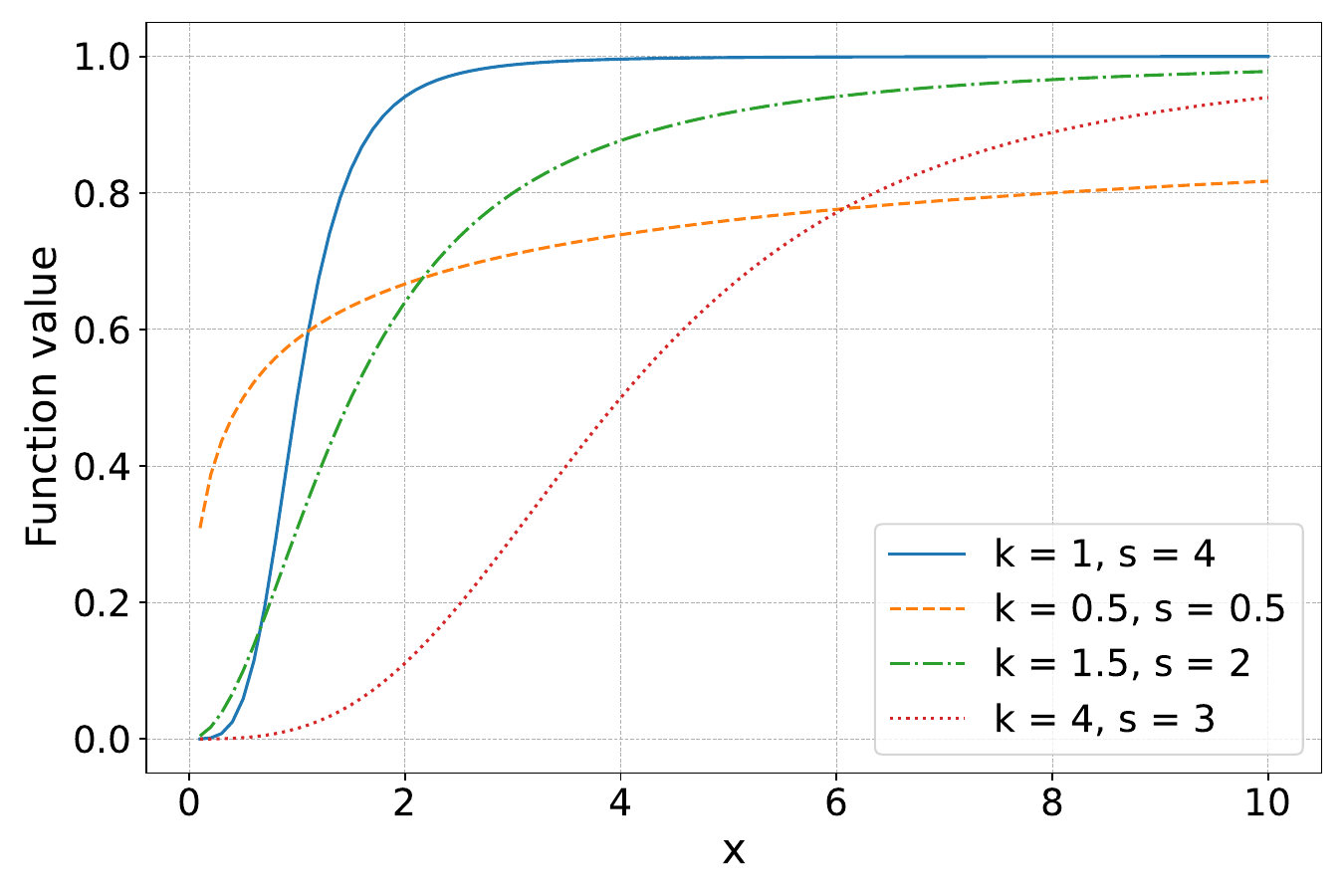}
    \tcaption{Examples of Hill functions}{Four examples of Hill functions with varying inflection point ($k$) and shape ($s$) parameters.}
    \label{fig:hill_examples}
\end{figure}

Modern practice has converged on two related functional forms to capture these types of effects. The first, more general form is the \textit{Hill} function, given by: 
\begin{equation}
    \mathrm{Hill}(x; k, s) = \frac{1}{1+\left( \frac{x}{k} \right)^{-s}}. \label{eq:hill}
\end{equation}
While modern MMM practice has adopted the term ``Hill'' for this function, based on its early use in pharmacology \citep{hill1910possible}, the functional form is equivalent to the well-known log-logistic CDF, and to the ADBUDG function introduced by \cite{little1970models}. This function has been used to model nonlinear effects in all recent MMM modeling libraries, including Google's Meridian \citep{meridian}, Meta's Robyn \citep{robyn}, and PyMC-Marketing \citep{pymc}. Examples of Hill functions with different parameters are shown in \Cref{fig:hill_examples}. While the Hill function is quite flexible, it also can be poorly identified, with different combinations of the parameters yielding effectively the same function, especially over a finite range \citep{jin2017bayesian}. Thus, it is sometimes simplified by setting $\mathcal{S} = 1$, yielding what \cite{jin2017bayesian} call the \textit{reach transformation}, so-called because of its use in modeling the relationship between reach and GRPs in TV advertising.

\subsection{Carryover Effects}

A second area of extensive research has been how to effectively model carryover effects of marketing. It has long been realized that marketing, and in particular, advertising, can have both short and long-run effects \citep{palda1965measurement}. Such an observation motivated the extensive development of time-series models for modeling ad effects, of the form,
\begin{equation} \label{eq:lagged_effects}
    y_{t} = \alpha + \beta_0 x_t + \beta_1 x_{t-1} + \beta_2 x_{t-2} + \ldots + \beta_L x_{t-L} + \varepsilon_t,
\end{equation}
where again, for simplicity, we focus on spend in single marketing channel, this time including multiple lags of that spending \citep[e.g.,][]{bass1972testing}. While such models are rich, they also introduce many parameters per channel, increasing both the data requirements, and risk of overfitting. Thus, to simplify the problem, the literature has turned to \textit{stock} variables, most notably the well-known \textit{AdStock} \citep{hanssens2003market,zantedeschi2017measuring}. These stock variables are weighted combinations of past spending,
\begin{equation}
    \mathrm{Stock}(x_{t}, \ldots, x_{t-L}) = \frac{\sum_{\ell=0}^L w_\ell x_{t-\ell}}{\sum_{\ell=0}^L w_\ell},
    \label{eq:adstock}
\end{equation}
where the weights are assumed to follow a particular known form, like geometric decay, given by:
\begin{equation}
    w^\mathrm{G}_\ell = \lambda^\ell, ~\lambda \in [0,1],
\end{equation}
or Pascal decay, given by the PMF of a Pascal (or negative binomial) distribution:
\begin{equation}
    w^\mathrm{P}_\ell = (1-\lambda)^\ell \lambda^\tau \left(\frac{\ell + \tau - 1}{\tau}\right), ~\lambda \in [0,1],
\end{equation}
which allows for the possibility of delayed peaks in marketing effectiveness \citep{bultez1979does}. In this way, stock variables can be used instead of the original variables in an MMM, and the MMM is thus interpreted as capturing the effects of changing the marketing stock on revenue, rather than specific instantaneous or lagged effects. In the case of infinite lags, with weights decaying geometrically, the model can be rewritten as the well-known Koyck model \citep{koyck1954distributed}, simplifying to:
\begin{equation}
    y_t = \alpha + \beta_0 x_t + \lambda y_{t-1} + \varepsilon_t.
\end{equation}

\subsection{Time-Varying Effects}

The idea that marketing effectiveness changes over time also has a long history in marketing \citep[e.g.,][]{parsons1975product, winer1979analysis, wildt1983modeling, naik1998planning, van2004dynamic, bruce2008pooling}. There are two types of models used in this literature: theory-based models, where nonstationarity in the effect of advertising is derived from an underlying theory of ad response, and unstructured models, where time-varying effects of advertising are estimated flexibly via time-dependent parameters in a reduced-form model, like regression. 

Seminal work in theory-based dynamic models includes \cite{naik1998planning}, who estimate a variant of the classical Nerlove-Arrow (NA) model of advertising response, where awareness from advertising is a function of advertising levels scaled by the quality of the ads, and past awareness scaled by a forgetting rate. \citeauthor{naik1998planning} extend this model to allow for time-varying quality of advertising, which thereby makes the effect of current period advertising also time-varying. Numerous other papers have built on and expanded the NA framework, including \cite{bass2007wearout}, who consider such dynamics across multiple advertising themes, and \cite{bruce2008pooling}, who adds an additional source of dynamics in the form of a time-dependent forgetting rate. Yet more work has considered similar dynamic models but in new advertising settings. Some examples include spillover effects in paid search ads \citep{rutz2011generic}, customer-firm relationship on Twitter \citep{ma2015squeaky}, the effects of format, content, and targeting in digital display advertising \citep{bruce2017dynamic}, and response of online brand search to TV ads \citep{du2019immediate}. While these papers allow for a rich, nuanced understanding of important forces in marketing, they also assume specific types of time-varying effects, and often assume specific parametric forms therein. For instance, \cite{naik1998planning} assume advertising quality decays over time, following a specific differential equation, while \cite{bruce2008pooling} assumes dynamic forgetting effects follow a generalized logistic transformation of advertising and goodwill. 

In contrast, as previously described, modern MMMs typically take the form of simple, regression-style models, where the goal is to estimate an elasticity, or, as outlined previously, a response curve. The simplest way to incorporate time-varying effects in such a model is to consider a standard linear (or log-linear) specification, and allow the coefficients to evolve with time:
\begin{equation}
    y_t = \beta_t x_t + \varepsilon_t. \label{eq:uni_dyn_model}
\end{equation}  
Such specifications have been used in the MMM setting by, e.g., \cite{winer1979analysis}, \cite{van2004dynamic}, \cite{mariel2005nonparametric}, and \cite{ng2021bayesian}. While the form of the model has remained consistent across these works, the actual specification for $\beta_t$ differs. As, in most cases, there is only one data point per time period, some amount of smoothing must be induced to enable estimation. For example, in \cite{winer1979analysis}, the coefficients are modeled following a random walk, while in \cite{van2004dynamic}, they follow a more sophisticated state-space model, incorporating independent variables in the evolution. In \cite{ng2021bayesian}, the evolution of the coefficients is modeled using a Bayesian dynamic factor model. 

While time-varying coefficient models are relatively rare in the aggregate data setting, they are more common in settings where several observations per period are available. Most notably, in the context of individual-level random utility models, there is a long history of modeling how individuals' sensitivities to the marketing mix change over time \citep[e.g.][]{kim2005modeling, liechty2005dynamic, sriram2006effects, lachaab2006modeling, guhl2018estimating, dew2020modeling, dew2024correlated}. Particularly relevant for the present work are \cite{dew2020modeling} and \cite{dew2024correlated}, who introduce Bayesian nonparametric specifications for the evolution of $\beta_t$ that are similar to what we propose subsequently. Models with time-varying parameters are also relevant in the context of causal inference, to measure how a causal effect is changing over time \citep[e.g.,][in the context of synthetic control]{klinenberg2023synthetic} or to allow for time-varying differences between cohorts \citep[e.g.,][in the context of an age-period-cohort analysis]{oblander2023frontiers}.

\subsection{Clarification on Terminology}

Before proceeding, we emphasize an important issue regarding terminology. Specifically, our focus throughout the rest of the paper will be on disentangling nonlinear effects (i.e., $y_t = f(x_t)$) and time-varying effects (i.e., $y_t = \beta_t x_t$). These effects are, as we just described, distinct from carryover effects. In both cases, carryover effects may be relevant: for instance, revenue may be assumed to be a nonlinear function of a stock variable (e.g., AdStock), thereby capturing potential nonlinearities like diminishing returns in both short and long-term ad effects. Similarly, time-varying models can also include carryover effects, by allowing the coefficient vector, $(\alpha, \beta_0, \beta_1, \ldots, \beta_L)$ to depend on time, $(\alpha_t, \beta_{0t}, \beta_{1t}, \ldots, \beta_{Lt})$. This time-varying formulation allows the analyst to capture effects like changes in the effectiveness of same-period advertising, or changes in the carryover effect of past advertising. 

Importantly, in the literature, the word ``dynamic'' has been applied to describe both models with carryover, including time series and state-space specifications, and models with time-varying coefficients, including dynamic factor models and dynamic heterogeneity specifications. Adding to the complexity of the terminology, the workhorse model of much of the literature on modeling carryover and persistence effects is the dynamic linear model (or linear state-space model), which, in its most general form, allows for both carryover-style effects \textit{and} time-varying effects \citep{pauwels2004modeling}. To avoid confusion throughout the remainder of the paper, we use ``dynamic'' interchangeably with ``time-varying'' to indicate that there are \textit{dynamics in the effectiveness of marketing}, though for clarity, we favor the latter. To avoid any confusion, for carryover effects of any type, including as captured by lagged independent variables, Koyck-like formulations, or stock variables, we will explicitly use the terminology ``carryover.'' 

\section{Theory} 

Given this background, we now return to the focal question: under what conditions will nonlinear and time-varying models be conflated, given simple MMM data? We assume, as before, that the two potential models we are interested in are either the nonlinear response model, $y_t = f(x_t) + \varepsilon_t$, or the time-varying linear model, $y_t = \beta_t x_t + \varepsilon_t$.\footnote{In reality, it is possible that both types of effects may coexist. In that case, these two models can be viewed as approximations, where the analyst's goal is to understand which effect is more dominant. The coexistence of the effects would further complicate identification issues.} To answer the focal question, we proceed in two parts: we first ask under what conditions the time-varying model can capture nonlinear effects, assuming the nonlinear model is the true data generating process (DGP), and then vice versa. 

\subsection{Time-varying Approximates Nonlinear}

We assume the true DGP is the nonlinear model,
$$ y_t = f(x_t) + \varepsilon_t, $$
with $\varepsilon_t \sim \mathcal{N}(0,\sigma^2)$, and with $x_t$ such that $\mathbb{E}(x_t \varepsilon_t) = 0$. Given a sequence $x_1, \ldots, x_T$, we use this model to generate data $(x_t, y_t), t=1, \ldots, T$, to which we then wish to fit the model, 
$$ y_t = \beta_t x_t + \epsilon_t. $$
As described in our review of time-varying models, given we have only one data point per period $t$, estimating such a model requires inducing some regularization on the coefficients, $\beta_t$. Thus, for now, we will assume that such a time-varying model is estimated by minimizing a regularized sum of squared errors,
\begin{equation}
\widehat{\beta}_1,\ldots,\widehat{\beta}_T=\underset{\beta_1,\ldots,\beta_T}{\arg\min}\left[\sum\limits^T_{t=1}(y_t-\beta_t\cdot x_t)^2+\Lambda(\beta_1,\ldots,\beta_T)\right]
\end{equation}
where the regularizer function $\Lambda(\beta_1,\ldots,\beta_t)$ enforces a relationship between the coefficients such that, a priori, coefficients of closer periods are expected to be more similar than those from distant periods. This framework captures many types of time-varying models, including extant specifications, as well as our proposed model, introduced in the next section.\footnote{Specifically, in our proposed framework, the prior on $\beta_t$, a Gaussian process, plays the role of this regularizer, with its lengthscale parameter governing the degree of coefficient similarity between adjacent periods. More generally, priors in Bayesian inference play the role of regularizers, as discussed in \citep[e.g.][]{dew2024probabilistic}.}

To understand to what degree this time-varying model will be able to capture the true DGP, we start with a few simple cases. First, let us examine what happens if there is no regularizer ($\Lambda(\beta_{1:T})\equiv 0$). In this case, the time-varying model can (hypothetically) achieve perfect fit, with $\widehat{\beta}_t = y_t/x_t$, regardless of the DGP. Of course, such a model will also fail to generalize, as there is no mechanism for relating $\widehat{\beta}_t$ across different $t$. Nonetheless, this demonstrates that the general time-varying linear framework is quite flexible.

On the other hand, let us consider an extreme amount of regularization:
\begin{equation}
    \Lambda(\beta_1,\ldots,\beta_T) =\begin{cases}
    0  & \text{if }\beta_1=\beta_2=\ldots =\beta_T\\
    \infty & \text{otherwise}.
    \end{cases}\label{eq:ols_regularizer}
\end{equation}
Such a regularizer forces all of the coefficients to be equal, which, under frequentist inference, reduces the problem to finding the OLS estimator for the data, which we denote as $\hat{b}$. The fit of such a model will thus depend on how well the nonlinear function can be approximated by a linear model. We can use a Taylor expansion to understand that approximation. If $f$ is at least twice differentiable, we can compute exactly the OLS estimator in terms of the original DGP primitives by using a Taylor expansion around the mean of $x$, denoted $\bar{x}$, as: 
\begin{align}
    \hat{b} &= \frac{1}{\sum_t(x_t-\bar{x})^2}\left[ f'(\xbar)\sum_t(x_t-\bar{x})^2+\frac{1}{2}\sum_tf''(\xi_t)(x_t-\xbar)^3+\sum_t(x_t-\xbar)\varepsilon_t\right]\nonumber\\
    &=  f'(\xbar)+\frac{1}{2}\frac{\frac{1}{T} \sum_tf''(\xi_t)(x_t-\xbar)^3}{\frac{1}{T} \sum_t(x_t-\bar{x})^2}+\frac{\frac{1}{T} \sum_t(x_t-\xbar)\varepsilon_t}{\frac{1}{T} \sum_t(x_t-\bar{x})^2},\label{eq:b_hat_as_fn_of_f}
\end{align}
where $\xi_t$ is a scalar between $x_t$ and $\bar{x}$. We include more details of this derivation in Web Appendix A. What this equation suggests is that the coefficient in our linear regression would be equal to the derivative of the response function $f$ in the DGP at $\bar{x}$, plus two error terms. The third term is zero in expectation, as we assume $\varepsilon$ and $x$ are independent. The second term depends on the second derivative of the response function. Thus, intuitively, the linear approximation works better when the function behaves ``smoothly,'' insofar as its second derivative is small. This result is neither new nor surprising: essentially, if the tangent line to $f$ captures $f$, the linear model can approximate $f$ very well. Instead, this derivation merely illustrates that the OLS result will be almost exactly equal to the derivative of the specific response function we care about. 

What is more interesting is to consider regularization schemes between the two extremes presented above. Intuitively, intermediate amounts of regularization should fall between the ``worst-case'' scenario of fitting as well as a static linear model, and the ``best-case'' scenario of actually \textit{overfitting} the data. An interesting specific example to consider is a piecewise linear regression with a fixed-size window $\tau$. To simplify the notation, we consider $T = N\cdot \tau$ such that we can partition the set $\lbrace 1,\ldots,T\rbrace$ into $N$ sets of size $\tau$, each set defined by $S_n=\lbrace (n-1)\cdot \tau+1, (n-1)\cdot \tau + 2, \ldots,n\cdot \tau  \rbrace$. The piecewise linear model is equivalent to a regularizer that enforces $\beta_t$ to be constant across periods within the same partition
\begin{equation}
    \Lambda(\beta_1,\ldots,\beta_T) =\prod^N_{n=1} \Lambda^{OLS}\left(\beta_{t\in S_n}\right)
\end{equation}
where $\Lambda^{OLS}$ is the regularizer from \Cref{eq:ols_regularizer}. In this case, the estimation problem is separable, and it suffices to run a separate OLS regression on each partition, with $\hat{b}_n$ the estimator for partition $n$. Interestingly, \Cref{eq:b_hat_as_fn_of_f} already provides an expression for each coefficient $\hat{b}_n$. Moreover, we can vary $N$ and $\tau$ and recover the previous two cases: if $N=1$ we have OLS, and if $\tau=1$ we have no regularization and perfect fit. This model is interesting not just because of its tractability, but because it is similar to the types of nonparametric models used in the recent literature for modeling time-varying effects \citep[e.g.,][]{dew2020modeling,ng2021bayesian}, insofar as those models inducing smoothing in $\beta_t$ over local regions of time. 

Putting these pieces together, our analysis suggests that even a moderately regularized time-varying linear model can approximate the original nonlinear DGP. The degree of accuracy of this approximation will depend on the smoothness of the original function. However, it also depends on one other important factor we have not yet discussed: how consistent $x_t$ is over $t$. From \Cref{eq:b_hat_as_fn_of_f}, if $x_t$ changes dramatically from period to period, and the true model is nonlinear, then we will have different values of $f'(\bar{x})$ in nearby time periods, and thus, very different estimated values of $\hat{b}_t$. Such changes will be incompatible with a smoothness assumption for $\beta_t$. Thus, a smooth $\beta_t$ will only be able to approximate the nonlinear DGP if $x_t$ does not vary too much over $t$, relative to changes in $f'(x)$. In practice, $x_t$ often does evolve slowly: for instance, spending decisions at firms are often made in an autocorrelated way (e.g., increasing spend by 10\%). Moreover, in cases where $x_t$ is a stock variable, smoothness in $x_t$ is induced by the stock variable itself. Thus, under many common regimes for setting or specifying $x_t$, the time-varying model can approximate the nonlinear model.

\subsection{Nonlinear Approximates Time-varying}

We now consider the opposite problem: given a time-varying linear data generating process, when could a nonlinear model approximate it? Mathematically, consider a true data generating process given (as before) by, 
\begin{align}
y_t &= \beta_t\cdot x_t+\varepsilon_t, \quad  \varepsilon_t\sim\N(0,\sigma^2).\label{eq:dynamic_dgp}
\end{align}
We want to understand under what conditions such a model can be equivalently expressed as a static nonlinear model, $y_t = f(x_t) + \epsilon_t$. 

While the argument for time-varying approximating nonlinear was fairly broad, the argument in this case is more narrow. First, let us further assume that the time-varying effectiveness, $\beta_t$, can be written as $B(t)$, a continuous function. As we rigorously show in Web Appendix A, the static nonlinear model can only approximate the time-varying linear model when $B(t)$ can be written as a function of spending, $x_t$. That is, if there is some way of writing $B(t) = h(x_t)$, for some function $h$. That this condition induces conflation is intuitive: if we can write $B(t) = h(x_t)$, then we can express the DGP as $y_t = h(x_t) x_t + \varepsilon$, which is equivalent to the nonlinear DGP with $f(x_t) = h(x_t) x_t$. This condition is, as we show in Web Appendix A, both necessary and sufficient for conflation. 

There are many practically relevant cases where this condition is satisfied. One such case is when $x_t$ is monotonic with respect to time (e.g., marketing spending always increases in the observation window). If spending is monotonic, then one can map each observed level of spending to the exact period it occurred. In turn, this means that each level of spending is associated with a specific level of marketing effectiveness, $\beta_t$, and therefore, it can be written as a function of $x_t$ (see Web Appendix A). While monotonicity is a strict requirement, it is not unrealistic: firms often loathe cutting marketing budgets, and thus, an (approximately) increasing trend in marketing spend on a channel is plausible.

Another case in which $B(t)$ can be expressed as a function of $x_t$ is if spending and effectiveness are both determined by a variable, $p_t$, and the relationship between this variable and spending is invertible. We call this variable $p_t$ a \textit{parent process}. For instance, consider the motivating example from the introduction: if seasonality determines marketing effectiveness, and spending on marketing is set following such seasonality, that seasonality is a parent process. In fact, the parent process can also be one of the two variables: for instance, even absent seasonality, if $x_t$ is set following $\beta_t$, $\beta_t$ can be viewed as the parent process. In such cases, we can write both $x_t$ and $\beta_t$ as functions of the parent process (i.e., $x_t=X(p_t)$ and $\beta_t=B(p_t)$). If the spending function is invertible, then for each level of observed spending, one can map that level to a corresponding level of the parent process, and then subsequently map that on to a corresponding marketing effectiveness, thus allowing us to write $B(t) = h(x_t)$ (i.e., $\beta_t=B(X^{-1}(x))$; for more details, see Web Appendix A).

In sum, if the true data generating process is a time-varying linear model, that same data can be captured by a static nonlinear model when the effectiveness of marketing, $\beta_t$, can be expressed as a function of $x_t$. In turn, this is likely to happen when either $x_t$ is monotonic with respect to time, or if $x_t$ and $\beta_t$ are both a function of some third variable. While these conditions may seem very specific, they are made more likely by common managerial practices: for instance, autoregressive spending decisions can make monotonicity more likely to occur. Another common practice is increasing (or decreasing) spending according to a ``flight path'' of a fixed percentage increase (or decrease) per period. Such flight paths would also induce monotonic spending. Setting spending following some knowledge about the evolution of effectiveness, or according to known seasonal variation, can also induce the problem. Finally, we note that, while the above theoretical results hold in a very strict sense, in practice, with noisy data, conflation may be more likely even if these conditions do not strictly hold. We will demonstrate this empirically in the following sections.

\section{Empirical Framework} 

To further explore the issue of conflation, and to bring these ideas to bear on real data, we need a flexible framework that can both simulate various nonlinear and time-varying effects, and estimate them from data. While the literature has proposed various models, as described in our literature review, they are often restrictive in the functional forms they assume, or do not cleanly allow for manipulating various features of the response functions that we demonstrated are important in our theoretical analysis, namely their smoothness. Thus, in this section, we introduce a general Bayesian nonparametric framework, based on Gaussian processes, that does so. 

To begin, we assume an additive regression model, of the form:
\begin{equation} 
    y_t = \alpha + g_1(x_{1t}, t) + g_2(x_{2t}, t) + \ldots + g_J(x_{Jt}, t) + \varepsilon_{t},
\end{equation}
where $x_{jt}$ represents spend on marketing channel $j$ at time $t$, $y_t$ is the outcome of interest, and $\varepsilon_t \sim \mathcal{N}(0,\sigma^2)$. Here, $y_t$ and the various $x_{jt}$ can be scaled in any meaningful way (e.g., log). We further assume this model simplifies into one of two forms: either the static nonlinear model, 
\begin{equation}
    y_t = \alpha + f_1(x_{1t}) + f_2(x_{2t}) + \ldots + f_J(x_{Jt}) + \varepsilon_{t}, \label{eq:mv_nl_model}
\end{equation}
or the time-varying linear model,
\begin{equation}
    y_t = \alpha + \beta_1(t) x_{1t} + \beta_2(t) x_{2t} + \ldots + \beta_J(t) x_{Jt} + \varepsilon_t. \label{eq:mv_dyn_model}
\end{equation}
In all three equations, the object of interest is a function: either a joint function, $g(x,t)$, which is often intractable to estimate; a function of spending, $f_j(x_{jt})$; or a function of time, $\beta_j(t)$. To generate and estimate such functions, we take a Bayesian perspective: given these unknown functions of interest, we assign them priors that capture uncertainty over functions. Bayesian approaches to MMM are the standard in industry \citep[e.g.,][]{jin2017bayesian}. In the literature on Bayesian nonparametrics, a flexible way of specifying priors for functions is through Gaussian processes \citep[GPs,][]{williams2006gaussian}. 

Recent years have seen a number of uses of GPs in marketing, including in capturing time-varying effects of the marketing mix \citep{dew2020modeling, dew2024correlated}. In brief, a GP is a stochastic process \(h(.)\), indexed by inputs $z \in \reals^D$. While the inputs can theoretically have any dimension, in our two focal cases, $D=1$, with $z = x_{jt}$ in \Cref{eq:mv_nl_model} or $z = t$ in \Cref{eq:mv_dyn_model}. Thus, in the following, we will assume $D=1$. A GP is defined by a \textit{mean function} \(m(z)\) and a \textit{covariance function} \(k(z,z')\), also called the \textit{kernel}, such that for a fixed set of inputs \(\boldsymbol{z} = \{z_1, z_2, \ldots, z_T\}\),
\[ 
h(\boldsymbol{z}) \sim \mathcal{N}(m(\boldsymbol{z}),k(\boldsymbol{z}, \boldsymbol{z}')),
\]
where \(m(\boldsymbol{z})\) is the mean function evaluated at all inputs, a \(T \times 1\) vector, and \(k(\boldsymbol{z}, \boldsymbol{z}')\) is a \(T \times T\) matrix formed by evaluating the covariance function \(k(z,z')\) pairwise at all of the inputs. Thus, for any given inputs, a GP defines a multivariate normal distribution over the associated function values, with relationships between them governed by the mean and covariance functions. The mean function encodes a prior expectation for $h$, while the covariance encodes how $h$ deviates from $m$, including crucial quantities like smoothness and amplitude. When used as a prior, as in our model, the GP is typically denoted by \(h(z) \sim \mathcal{GP}(m(z), k(z,z'))\).

\begin{figure}
    \centering
    \includegraphics[width=0.8\textwidth]{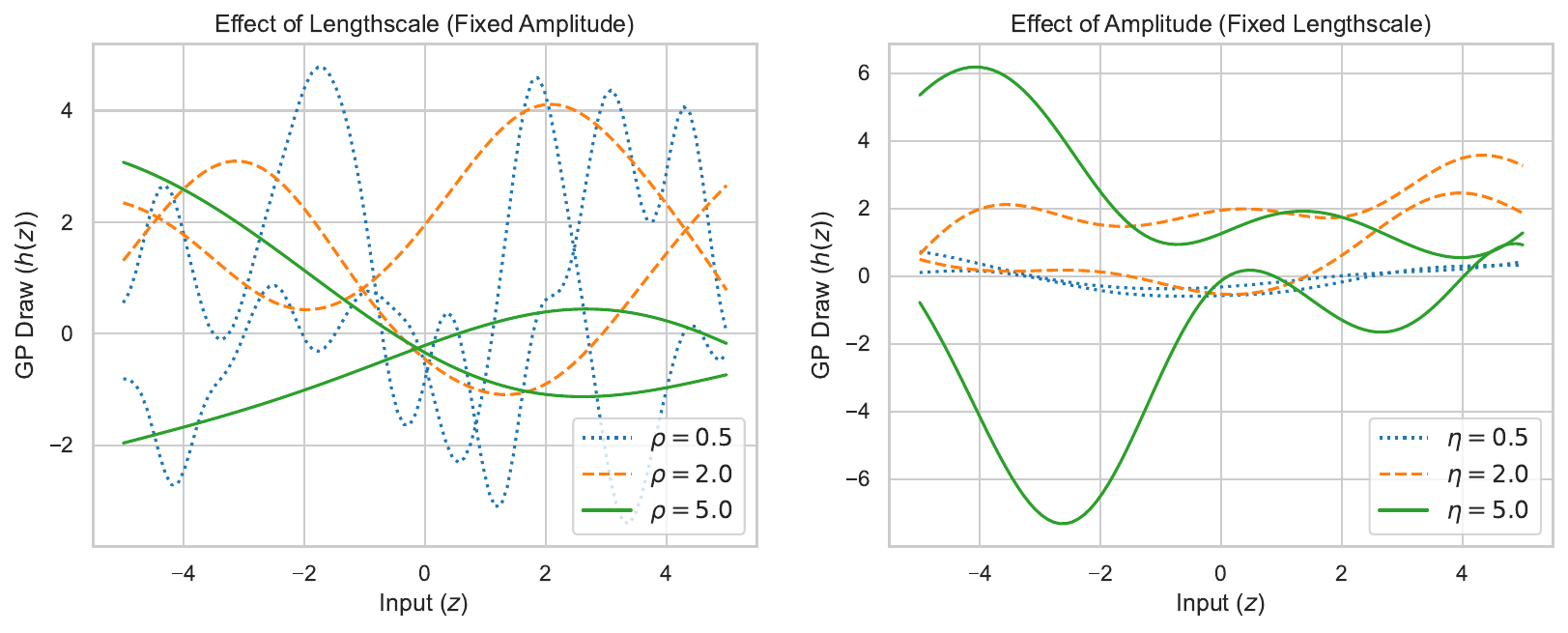}
    \tcaption{Examples of GPs}{At left, six draws from GPs, varying the lengthscale parameter $\rho$ while keeping the amplitude fixed at $\eta=2$. At right, six draws from GPs, varying the amplitude $\eta$ while keeping the lengthscale fixed at $\rho=2$.}
    \label{fig:gp_examples}
\end{figure}

We adopt GPs for three main reasons. First, GPs are a natural prior for unknown functions \citep{williams2006gaussian}, as in the nonlinear model, and have also seen successful use in capturing parametric evolution in time-varying models \citep{dew2020modeling}. Second, GPs are flexible: while the mean function $m(z)$ can be used to encode prior expectations about the shape of the function of interest, the kernel allows for nonparametric deviations from that expectation. Third, based on assumptions about the kernel, we can both generate and estimate functions with different properties, which is crucial for demonstrating the conflation issue. In particular, the kernel we adopt in this work is the simple squared exponential (SE) kernel, given by:
\begin{equation}
    k_\mathrm{SE}(z,z'; \eta, \rho) = \eta^2 \exp\left(-\frac{(z-z')^2}{2\rho^2}\right).
\end{equation}
Intuitively, the SE kernel captures the idea that nearby inputs, in terms of $(z-z')^2$, should have similar outputs, $h(z), h(z')$, with the level of similarity determined by $\rho$. $\rho$ is a hyperparameter of the kernel, referred to interchangeably as the \textit{lengthscale} or \textit{smoothness}. The other hyperparameter, $\eta$, controls the variance when $z=z'$, and thus determines the overall variability or \textit{amplitude} of $h(z)$. We demonstrate how these hyperparameters lead to different functions generated from the GP in \Cref{fig:gp_examples}.

Returning to our focal models, both can be simulated from, and estimated using GPs. Specifically, for the nonlinear model, for each $j$, we assume,
\begin{equation}
    f_j(x) \sim \mathcal{GP}(0, k_\mathrm{SE}(x,x'; \eta_j, \rho_j)).
\end{equation}
For the time-varying model, for each $j$, we assume,
\begin{equation}
    \beta_j(t) \sim \mathcal{GP}(0, k_\mathrm{SE}(t,t'; \eta_j, \rho_j)).
\end{equation}
That is, in both cases, we assume the unknown function comes from a GP with zero mean, and a squared exponential kernel. The zero mean assumption is the most general, as it imposes no prior knowledge. This could be further generalized to allow for, e.g., using the Hill function as the mean function in the nonlinear case. Note also a slight abuse of notation: while we have used the same notation for the hyperparameters across the two models, they need not be the same.\footnote{In fact, the priors used for these parameters will depend somewhat on the data and how it is normalized, as we will discuss more in our applications and in Web Appendix B.}

\paragraph{Alternatives to GPs} 

Before proceeding, we briefly describe a few alternatives to GPs that could be applied in this context. For modeling unknown functions, as in the nonlinear model, splines are a natural alternative \citep[e.g.,][]{kim2007capturing,brezger2008monotonic,boughanmi2021dynamics}. Notably, some variants of splines, like the P-splines of \cite{brezger2008monotonic}, allow for the imposition of monotone effects, which can be useful for MMM. Another potential alternative would be capturing nonlinear effects through a generalized additive model \citep{hastie1987generalized}. For capturing dynamics, including carryover and time-varying effects, time series and state-space models are a common tool  \citep{pauwels2004modeling}, as discussed previously. While more commonly used for specifying carryover effects, one could also model the evolution of $\beta_t$ in such a way. Such an assumption is used, for instance, in \cite{van2004dynamic}, who do so in a marketing mix setting, and \cite{lachaab2006modeling}, who use a state-space formulation for the evolution of coefficients in a choice model. While we favor GPs as a parsimonious solution that can capture both nonlinearities and time-varying effects, our central research question --- i.e., the conflation of these effects --- is orthogonal to this modeling choice. We demonstrate a few of these alternatives in the MMM context in Web Appendix C. Finally, we note that there are also deep theoretical connections between GPs and specific types of state-space models \citep{loper2021general}.

\section{Simulations}

With an empirical framework in hand for both simulating and estimating nonlinear and time-varying MMMs, we now return to the question of when the two will be conflated. While our theory offers some insights, it also relies on assumptions that may not strictly map onto the empirical framework described above. Thus, in this section, we explore the issue of conflation through simulation. We first conduct a ``mega-simulation'' to understand when conflation is a problem. Then, through smaller and more focused simulations, we explore the implications of conflation for marketing practice.

\subsection{Mega-Simulation}

Recall that our theory suggests the following: first, a time-varying model can approximate a nonlinear one when the nonlinearity is sufficiently smooth. Second, a (static) nonlinear model can approximate a time-varying one only when the dynamic marketing effectiveness can be expressed as a function of spending. While this condition is, theoretically, strict, we noted it maps onto several real-world conditions, including monotonic spending regimes, and cases where effectiveness and spending are jointly determined. Finally, in both cases, we noted that limited variation in or specific patterns of spending, as in autocorrelated spending decisions or adstocked spending, may exacerbate the issue. While this theory gives us some sense that conflation may be an issue, to understand when and to what degree it may actually materialize in practice, we ran an extensive set of simulations. 

\paragraph{Simulation Design}

We used three data generating processes: (1) the nonlinear model from \Cref{eq:mv_nl_model}; (2) the time-varying model from \Cref{eq:mv_dyn_model}; and (3) the nonlinear model but using specifically the Hill nonlinearity from \Cref{eq:hill}, rather than a GP. In each case, we assumed a single spending channel, $x_t$, which was also generated stochastically. Across the three simulations, we varied six parameters each across three levels (low, medium, and high):
\begin{itemize}
    \item \textbf{Properties of the Functions:}
    \begin{enumerate}
        \item[(a)] For the GP-based models, we varied the amplitude $\eta$ and smoothness $\rho$ parameters of the GP's kernel, effectively changing how much the effect can vary (through $\eta$) and how smooth it is (through $\rho$). For $\eta$, we set the low value to 1, the medium value to 2, and the high value to 5. For $\rho$, we set its value as a ratio of the range of the independent variable ($x_t$ for the nonlinear model and $t$ for the time-varying model). Namely, we set this ratio to be 0.1 (low), 0.5 (medium), and 1 (high).
        \item[(b)] For the Hill-based DGP, we varied the two Hill parameters: shape $s$ and inflection point $k$. We let the shape parameter take values 0.5, 2, and 3.5, while the inflection point was, similar to $\rho$, set as a ratio of the range of $x_t$, taking values 0.1, 0.33, and 1.
    \end{enumerate}
    
    \item \textbf{Properties of Spending:} We generated $x_t$ to follow a simple AR(1) autocorrelation process: $x_t \sim \mathcal{N}(\gamma_0 + \gamma_1 x_{t-1}, \tau^2)$, and varied $\gamma_1$ between 0 (no autocorrelation), 0.5 (medium autocorrelation) and 1 (random walk). In addition, we varied the transition variance $\tau$ to take values of 1, 5, and 10.

    \item \textbf{Noise in the Response:} We varied the noise in $y_t$ (i.e., the standard deviation of the error term in the DGP) as a ratio of its deterministic component, setting the ratio to the values 0.01, 0.1, and 0.2. 

    \item \textbf{Level of Carryover:} Finally, we varied the level of carryover in an AdStock formulation for $x_t$, from 0 (no carryover), to 0.3 (modest carryover) to 0.8 (high carryover).
\end{itemize}
For each of our 2,187 settings (i.e., 3 DGPs, each with 6 parameters with 3 settings each), we generated 100 different datasets, assuming 100 periods of observation. We then estimated both the nonlinear and the time-varying GP models. 

To test whether there is conflation, for each simulation, we held out a series of 10 observations, and then evaluated the model's mean squared error (MSE) on predicting revenue in those holdout periods. We labeled a simulation as conflated if the incorrect model, meaning \textit{not} the true DGP, performs equivalently to or better than the model with the true DGP by this metric. Note that this measure of conflation is an extremely high bar: we are asking that the competing model perform \textit{on par or better than the true DGP}. Ultimately, our main metric of interest is the rate of conflation for a specific simulation setting, or the percentage of simulations that were conflated.

\begin{figure}
    \centering
    \adjustbox{width=1.1\textwidth,center}{
    \includegraphics[width = 0.3\textwidth]{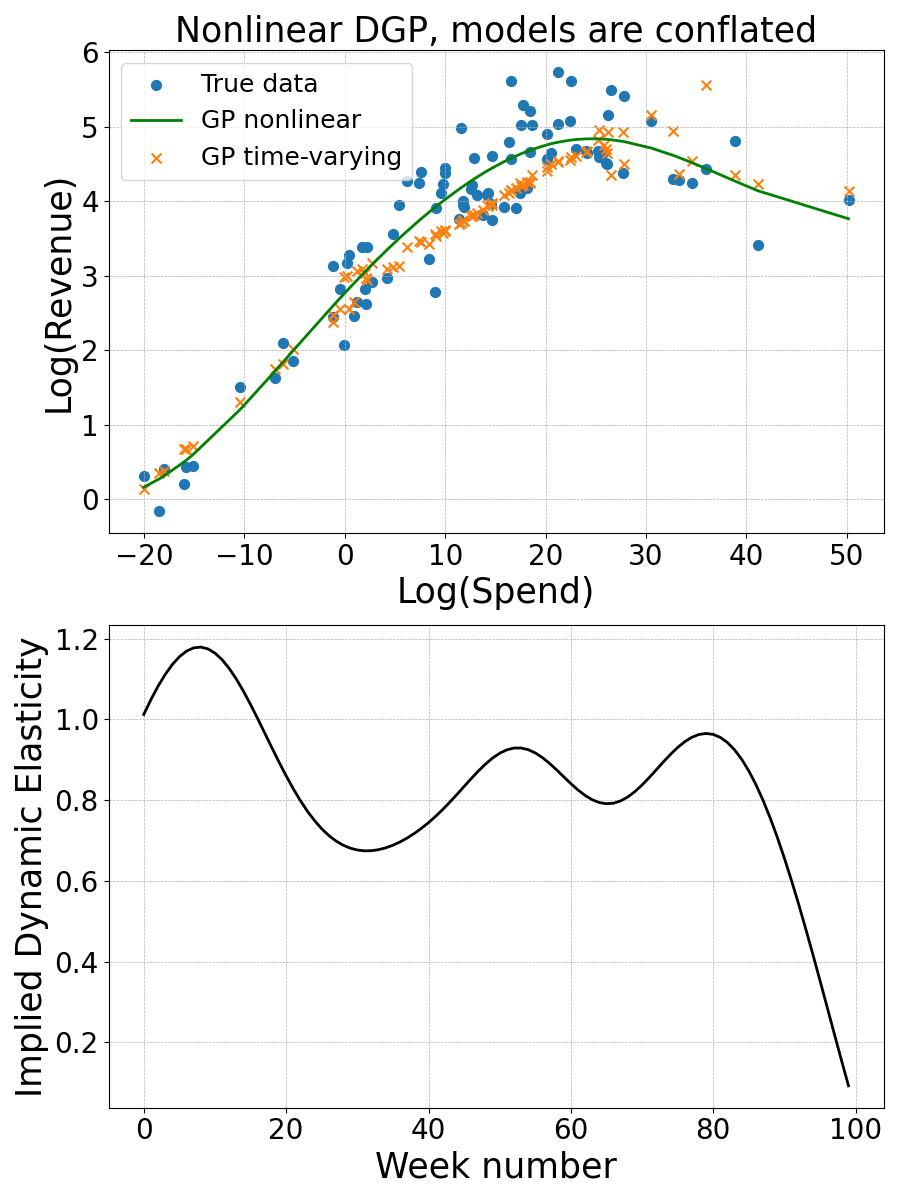}
    \includegraphics[width = 0.3\textwidth]{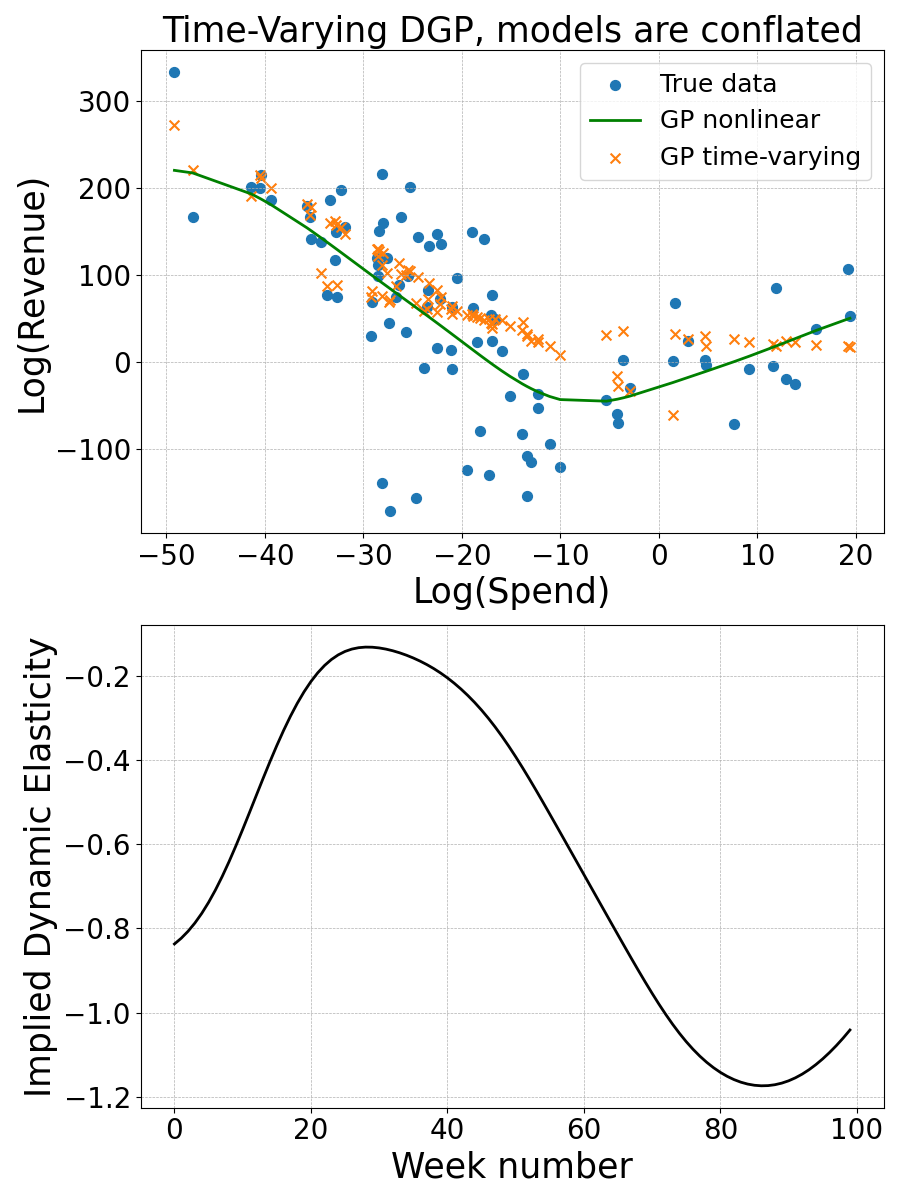}
    \includegraphics[width = 0.3\textwidth]{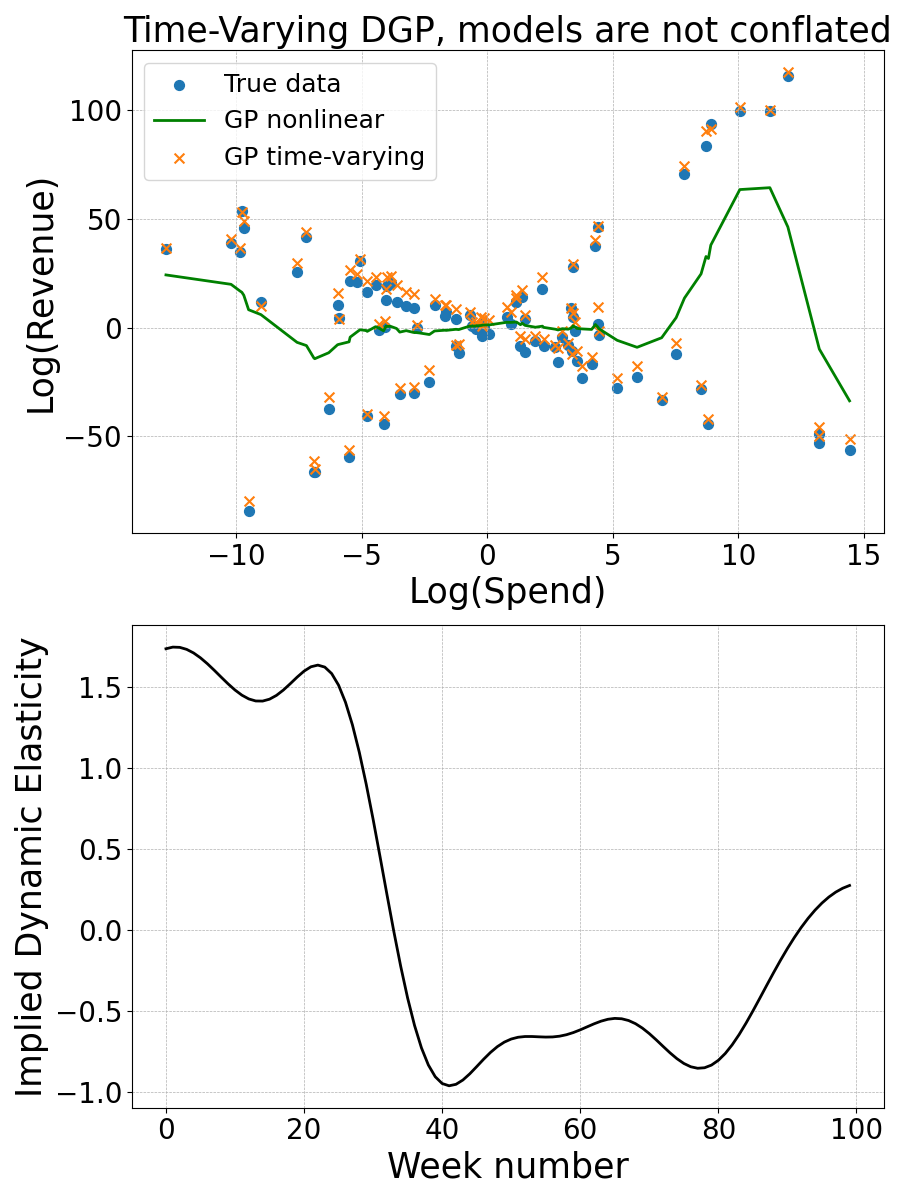}
    }
    \tcaption{Examples of Simulated Datasets and Conflation}{From left to right, simulated datasets where (1) the truth is nonlinear, but is conflated with time-varying; (2) the truth is time-varying, but is conflated with nonlinear; (3) the truth is time-varying, and is not conflated with nonlinear. At top are the true (simulated) data and the predicted values from each model; at bottom are the implied $\beta(t)$ from estimating the time-varying model.}
    \label{fig:sim_examples}
\end{figure}

To illustrate how these simulations work, we show three examples in \Cref{fig:sim_examples}. The examples are (1) a nonlinear DGP with a moderately smooth $f$; (2) a time-varying DGP with a moderately smooth $\beta(t)$; and (3) a time-varying DGP with a less smooth $\beta(t)$.\footnote{The exact specifications are: (1) $\eta = 5, \rho = 0.5, \gamma_1 = 1, \tau = 5, \sigma =  0.1, \alpha = 0$; (2) $\eta = 5, \rho = 0.5, \gamma_1 = 1, \tau = 5, \sigma =  0.1, \alpha = 0$; (3)~$\eta = 5, \rho = 0.1, \gamma_1 = 0, \tau = 5, \sigma =  0.01, \alpha = 0$.} In the top row of plots, we show the true data in blue, the predicted values from the nonlinear model in green, and the predicted values from the time-varying model in orange. In the bottom row, we plot the estimated $\beta(t)$ from the time-varying model. First, focusing on the top row, we see the first two examples are conflated: while we do not show the actual out-of-sample MSEs, we see that both the nonlinear fit (i.e., the line) and the time-varying fit (i.e., the xs) are overlapping, and almost perfectly fit the data. On the other hand, in the third case, while the time-varying model fits perfectly, the nonlinear model cannot approximate the data, which appears almost heteroscedastic due to the way that $\beta(t)$ is changing.\footnote{In fact, we observed that these heteroscedastic-looking patterns appear quite often when the DGP is time-varying and non-smooth. This is due to $\beta_t$ taking drastically different values across time, resulting in $y_t$ being essentially a mix of a few linear dependencies on $x_t$.} Moving to the bottom row, we see that the implied $\beta(t)$ in all three cases is relatively smooth, and appears plausible, even when the true DGP is nonlinear. That is, there are no material differences in the nature of the estimated $\beta(t)$ between a true time-varying model and a true nonlinear model. 

\paragraph{Results} Overall, we find widespread conflation. We summarize the results in \Cref{tab:sim_res_summary}, breaking the results down by the value of the carryover parameter $\alpha$ and focusing on the cases of no carryover ($\alpha = 0$) and high carryover ($\alpha = 0.8$)\footnote{$\alpha = 0.3$ results are very similar to $\alpha = 0$, as illustrated in \Cref{tab:sim_deep_dive}.}. In the case where there is no stock variable, under all DGPs, the vast majority of the simulation settings exhibited some conflation, with a sizable portion --- from 27\% to 46\% --- exhibiting major conflation, which we defined as over a quarter of that setting's simulations being conflated. Moreover, as predicted by our theory, these numbers only worsen when $x_t$ is modeled with high carryover: in that case, roughly 90\% of all simulation settings exhibited non-zero conflation, and the rate of major conflation also increased for all DGPs. These results suggest that, under a flexible data generating process, conflation will frequently occur. They also confirm that the smoothness of $x_t$ over $t$, especially as induced by stocking $x_t$, exacerbates the issue.

\begin{table}[p]
    \centering
    \begin{tabular}{lcccccc}
        \toprule
        & & \multicolumn{2}{c}{No Stock} & & \multicolumn{2}{c}{High Stock} \\
        \cmidrule{3-4} \cmidrule{6-7}
        True DGP & & Any &  Major & &  Any & Major \\
        \midrule
        Nonlinear & & 81\% & 27\% & & 85\% & 30\% \\
        Time-varying & & 91\% & 40\% & & 99\% & 44\% \\
        Hill & & 83\%  & 46\% & & 92\% & 47\% \\
        \bottomrule
    \end{tabular}
    \tcaption{Summary of Conflation}{For each DGP, the column ``Any'' reports the percentage of simulation settings that yielded a nonzero rate of conflation (i.e., any conflation). The column ``Major'' reports the percentage that yielded more than 25\% conflation.}
    \label{tab:sim_res_summary}
\end{table}

\begin{table}[p]
\centering
\resizebox{0.7\textwidth}{!}{%
\begin{tabular}{llrrrrr}
\toprule
 & & \multicolumn{2}{c}{{Nonlinear DGP}} & &\multicolumn{2}{c}{{Time-varying DGP}} \\
\cmidrule(lr){3-4} \cmidrule(l){6-7}
                {Variable} & {Level}  & {Coef} &  {$P(>\!|t|)$}    & & {Coef} & {$P(>\!|t|)$} \\
\midrule
Amplitude, $f$: & Low (1)     & - & - & & - & \\
 & Middle (2) & -0.24         & 0.65        &                  & 0.23          & 0.62                       \\
&   High (5)  & -0.65         & 0.22        &                  & 0.09          & 0.85                       \\
\midrule
Smoothness, $f$: & Low (0.1) & - & - & & - &  \\
 & Middle (0.5) & 10.36          & 0.00       &                   & 7.93          & 0.00                       \\
 & High (1)  & 18.86         & 0.00       &                   & 12.21          & 0.00                       \\
 \midrule
AR coef, $x$: & Low (0) & - & - & & - &  \\ 
 & Middle (0.5)           & -0.56          & 0.29        &                  & 0.63          & 0.18                       \\
& High (1)           & 0.42          & 0.43          &                & 13.38          & 0.00                       \\
\midrule
AR Variance, $x$: & Low (1) & - & - & & - &  \\
& Middle (5)        & 0.11         & 0.84           &               & 0.06          & 0.90                       \\
& High (10)       & -0.47          & 0.37               &           & 0.36         & 0.44                       \\
\midrule
Noise, $y$ & Low (0.01) & - & - & & - &  \\
& Middle (0.1)      & 14.10          & 0.00             &             & 13.05          & 0.00                       \\
& High (0.2)     & 23.28         & 0.00              &            & 22.71         & 0.00                       \\
\midrule
AdStock & Low (0) & - & - & & - &  \\
& Middle (0.3)      & -0.14          & 0.79             &             & 0.83          & 0.08                       \\
& High (0.8)     & 1.59         & 0.00              &            & 2.49         & 0.00                       \\
\bottomrule
\end{tabular}}
\tcaption{Simulation Results}{DV = Percentage Conflation; Intercept omitted for clarity.}
\label{tab:sim_deep_dive}
\end{table}

To develop a more nuanced understanding of when conflation is likely to be problematic, we further explored the results through regression analysis, this time focusing just on the nonlinear and time-varying DGPs. Specifically, we took the rate of conflation for a specific simulation setting as the dependent variable, and regressed it on the simulation's settings, coded as dummy variables. The results are displayed in \Cref{tab:sim_deep_dive}. In this model, positive coefficients indicate a simulation was more likely to be conflated if it had that feature. We see that the results are fairly consistent with our theory: for instance, high smoothness in $f$ is associated with much higher rates of conflation under both DGPs. In both cases, there is also a positive effect of the high AdStock coefficient. Interestingly, for the time-varying DGP, the AR coefficient in $x_t$ is highly significant, while it is not so for the nonlinear DGP. This asymmetry confirms our theory-based hypothesis that the conditions under which conflation arises are different for different DGPs. Finally, we see that conflation is dramatically more likely when the data are noisier. This last point is somewhat obvious, but worth mentioning: when data are noisy, it becomes harder to tell which model is right. As we will show in our application, modern MMM data \textit{are} often very noisy, meaning that again, conflation is highly likely in practice.

\subsection{Implications for Budget Allocation}

Until now, we have focused exclusively on documenting the potential \textit{existence} of conflation. We now turn to understanding its implications, again drawing on simulations. Recall that our definition of conflation is that the two models predict holdout data identically, such that an analyst could not choose the correct model by standard model selection metrics. Such holdout analyses are the standard way of doing model selection in practice. By definition, the holdout data used in such an analysis is generated under the status quo of spending as usual. Yet, just because two models give equivalent predictions under status quo spending patterns does \textit{not} imply that they will give the same predictions \textit{under intervention}. The distinction between status quo and intervention is important because optimizing an advertising budget is inherently interventional. Thus, time-varying models and nonlinear models may yield different recommendations for optimal budget allocation. 

To better understand this point, we first outline how to optimize a marketing budget, given an MMM. For simplicity, we focus here on the case when there are no carryover effects, and return to the more general case in our later empirical application. The goal of optimal budget allocation is to decide a level of spending on marketing channels, $x_{1t}, \ldots, x_{Jt}$, that maximizes expected profits. If there is no carryover effect of advertising, then optimizing spend is relatively straightforward. To do so, we consider the expected profits in a test period $t$, as predicted by our MMM, then maximize over $x_{jt}$:
\begin{equation}
    \max_{\substack{x_{1t}, \ldots, x_{Jt}}} \hat{y}_t(x_{1t}, \ldots, x_{Jt}) - \sum_{j} x_{jt}, \label{eq:spend_optim_no_carryover}
\end{equation}
where $\hat{y}_t(x_{1t}, \ldots, x_{Jt})$ is the predicted revenue from the MMM under a given level of spending. For the nonlinear model, this optimization can only be done numerically. For the time-varying model, the optimization can be done either numerically or, in some cases, in closed form. Assuming, for instance, a time-varying model with logged inputs and a single marketing channel, $x$, the optimal level of spending is:
\begin{equation}
x^* = \left[\exp(\alpha)\cdot \beta(t)\right]^{\left[1-\beta(t)\right]^{-1}}, \label{eq:closed_form_spend_optim_dyn}
\end{equation}
where $\alpha$ is the intercept of the log-log model. The result for a multichannel log-log model is analogous.

Now, we examine how model conflation relates to these budget allocation practices. Consider the approximation mechanisms described in our theory: the time-varying model approximates the nonlinear by effectively making a local linear approximation around $x_t$. Learning a smooth $\beta_t$, meaning a predictable $\beta_t$, requires reasonably smooth changes in $x_t$. In the optimal budget allocation procedures, we optimize expected revenue by intervening on $x_{t+1}$, thus breaking that smoothness: under these counterfactual spending levels, the predicted $\beta_{t+1}$ may no longer be a reasonable linear approximation to $f(x_{t+1})$. In the other direction, intervening on $x_t$ can break any relationship that exists between $x_t$ and $\beta_t$, thus again breaking the ability of the nonlinear to approximate the time-varying. From a predictive perspective, this means that the value of $\beta_{t+1} x_{t+1}$ may be very different from the (incorrect) implied $f(x_{t+1})$. 

\begin{figure}
    \centering
    \includegraphics[width=0.4\textwidth]{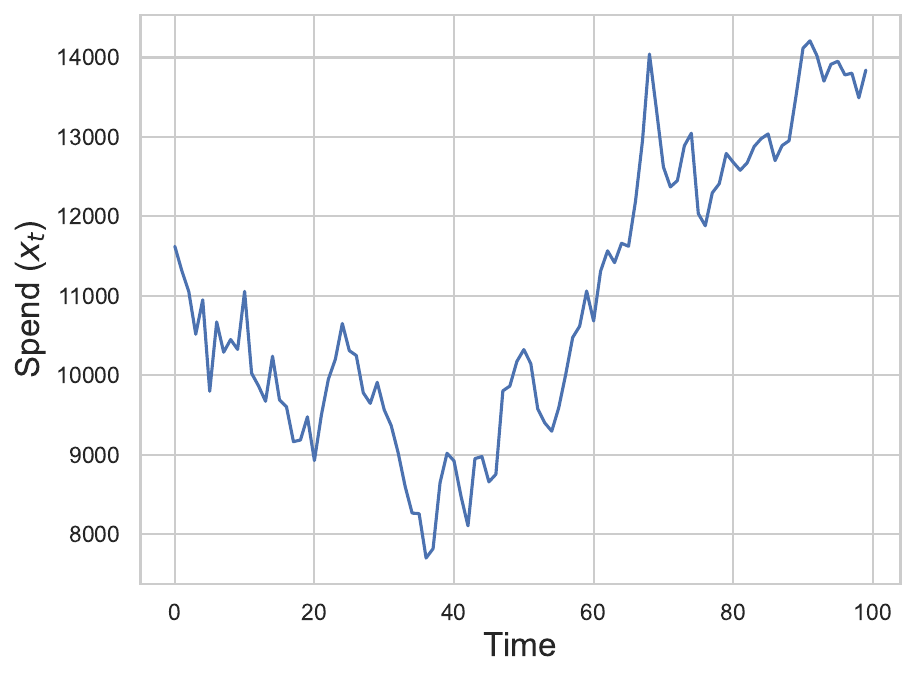} \includegraphics[width=0.4\textwidth]{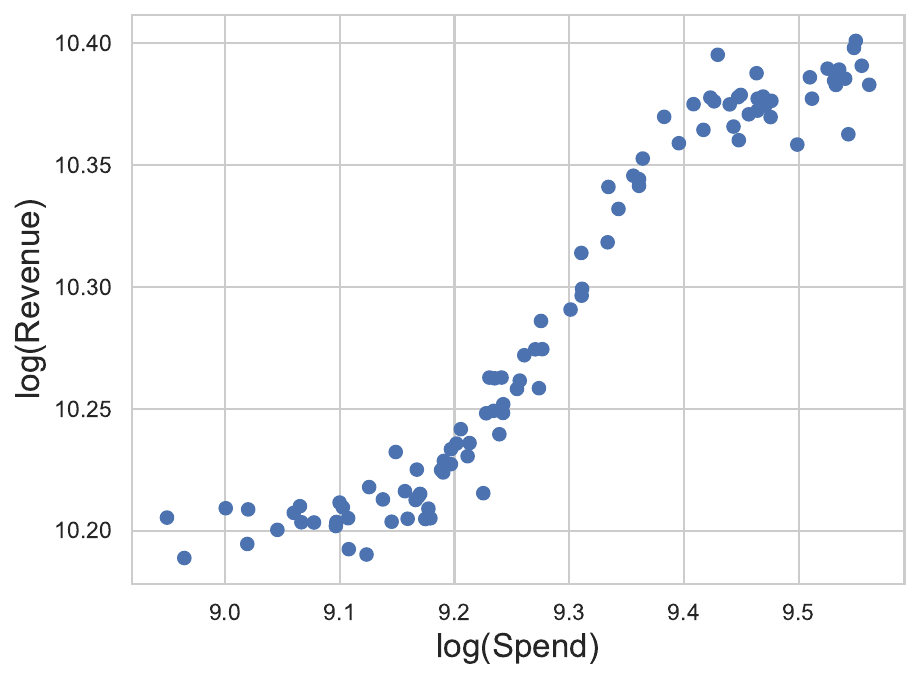}
    \tcaption{Spend Optimization Example, Data}{At left, the simulated evolution of spending, $x_t$, following an autoregressive process. At right, the simulated revenue, following a sigmoid (nonlinear) relationship.}
    \label{fig:sim_optim_data}
\end{figure}

\begin{figure}
    \centering
    \includegraphics[width=0.4\textwidth]{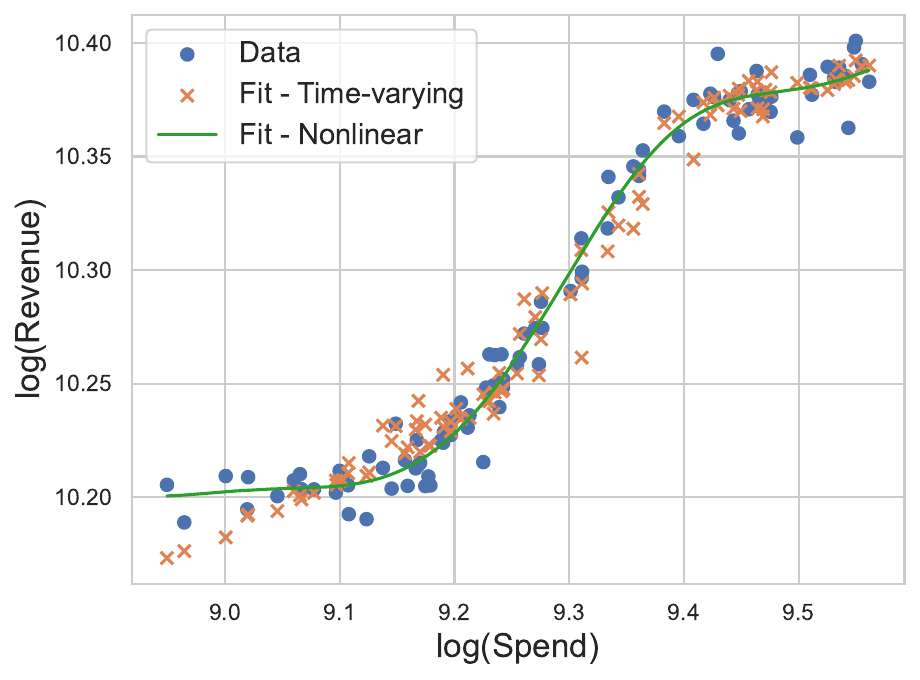} \includegraphics[width=0.4\textwidth]{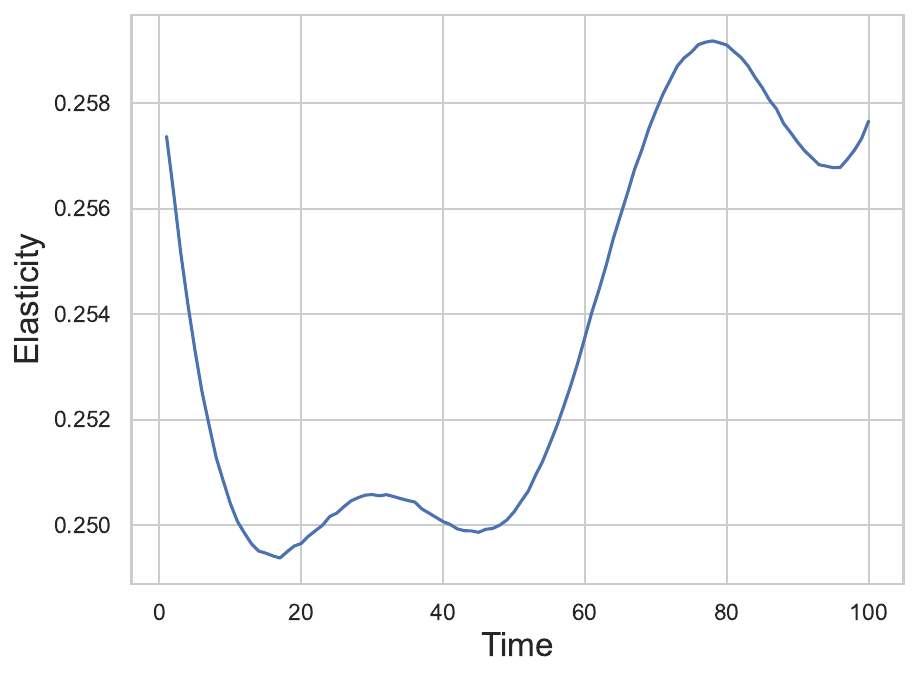}
    \tcaption{Spend Optimization Example, Model Fits}{At left, fit of both the nonlinear and time-varying models on the simulated data. At right, the estimated evolution of $\beta(t)$ recovered by the time-varying model.}
    \label{fig:sim_optim_fit}
\end{figure}

To illustrate this point empirically, we again turn to simulation. Consider the simulated dataset shown in \Cref{fig:sim_optim_data}. In this dataset, spending was generated following an autoregressive process, with a high degree of autocorrelation, similar to the mega-simulation, as shown in the left panel of the figure. The true DGP is nonlinear, following, in this case, a pre-defined sigmoid, with a low level of noise, as shown in the right panel. Here, we assume that the model is estimated on logged spending and revenue, with no carryover effects. With this data, we estimated both the nonlinear and time-varying models. They are, as expected, conflated, with the fits of both models shown in the left panel of \Cref{fig:sim_optim_fit}. We also show the implied elasticity over time, in the right panel of \Cref{fig:sim_optim_fit}. 

Now, given each model, we turn to the optimal budget allocation procedures described previously: for the nonlinear model, we optimize revenue by simple enumeration, tracing out a grid of possible expenditures within the range of the training data, then computing the expected profit at each point. In the time-varying case, we use the predicted coefficient $\beta_{T+1}$, and evaluate optimal expenditure using the closed-form expression given in \Cref{eq:closed_form_spend_optim_dyn}. 

The optimal spending across the two models is quite different: the optimal expenditure implied by the nonlinear model is \$12,072.60, while the optimal expenditure implied by the time-varying model is \$7,451.50, a difference that amounts to almost the entire range of the training $x_t$. These differences in recommended spending are driven by the different predictions made by the model in terms of how revenue would be affected by changing spending in time $T+1$: the time-varying linear model is assuming a log-log relationship with elasticity $\beta(T+1)$. On the other hand, the static nonlinear model assumes the revenue response is given by the curve shown in Figure \ref{fig:sim_optim_fit}, which is \textit{not} a log-log relationship. Under the interventional query of optimal spending, the nonlinear model suggests that it is worthwhile to invest enough money to get past the inflection point of the sigmoid. On the other hand, the log-log time-varying model has no such inflection point, and thus suggests a much lower optimal spend, given the implied elasticity at that point in time.

\section{Applications} 

Having identified both the potential for and implications of conflation in synthetic data, we now turn to several applications using real MMM data. We focus on two types of datasets: first, we revisit two classic datasets from the advertising response literature --- the dietary weight control product data of \cite{bass1972testing}, and the Lydia Pinkham data of \cite{palda1965measurement} --- to understand whether even in simple, storied datasets, the conflation issue arises. Second, we assemble a multitude of modern MMM datasets, using advertising data from Nielsen Ad Intel, to show how the issue may manifest in modern MMM practice.

\subsection{Classic Advertising Datasets}

In this section, we conduct two extensive analyses of classic advertising datasets: three years of sales and advertising data for a dietary weight control product (DWC), as analyzed in \cite{bass1972testing}; and the seasonally-adjusted Lydia Pinkham (LP) data, which cover 78 months of sales of Lydia Pinkham's herbal medication from January 1954 to June 1960, as analyzed in \cite{palda1965measurement}, \cite{winer1979analysis}, and many other studies. Both datasets have the same form, with just two variables --- sales and advertising --- defined on a monthly basis. The only difference is that, for DWC, sales are in units, while in LP, sales are in dollars. We analyzed each dataset with the same procedure: first, we applied a simple stock transformation to advertising, drawing on results in the literature to set the carryover amount ($\lambda$) and lag time ($L$).\footnote{Fixing these values, rather than estimating them, is common practice \citep[e.g.,][]{shapiro2021tv}. One reason is that, as \cite{jin2017bayesian} note, these parameters are often very poorly identified, and, under Bayesian implementations, strongly influenced by the prior.} Next, we fit four models: nonlinear (GP), time-varying (again, our GP-based version), nonlinear with the reach variant of the Hill function, and the time-varying model with a log-transformed stock variable. Finally, we evaluated their performance out-of-sample, evaluating conflation as in the simulations, and assessed the implications of each model for optimizing spending in the following period. We present the combined results on conflation in \Cref{tab:dwc_lp_rmse}, and discuss each case in more detail below. 

\begin{table}[h]
    \centering
    \tcaption{Classic Datasets --- RMSE Comparisons}{Each statistic is the posterior mean RMSE for each model. Below in brackets are the 95\% posterior intervals.}
    \label{tab:dwc_lp_rmse}
    \resizebox{0.9\textwidth}{!}{%
    \begin{tabular}{lcccc}
    \toprule
                          & Nonlinear         & Time-varying           & Hill & Log Time-varying               \\ \midrule
    Dietary Weight Control & 3.23                 & 2.97                 & 3.10 & 2.55                         \\
                          & [2.22, 5.58]          & [1.72, 5.09]         & [2.59, 4.44] & [1.50, 4.13]             \\ \midrule
    Lydia Pinkham         & 89.37                 & 83.21                & 100.50 & 88.92                       \\
                      & [79.41, 104.88]       & [75.44, 95.24]        & [91.68, 114.97] & [83.23, 98.00]        \\
    \bottomrule
    \end{tabular}}
\end{table}

\paragraph{Dietary Weight Control (DWC)}

\begin{figure}
    \centering
    \includegraphics[width=0.4\textwidth]{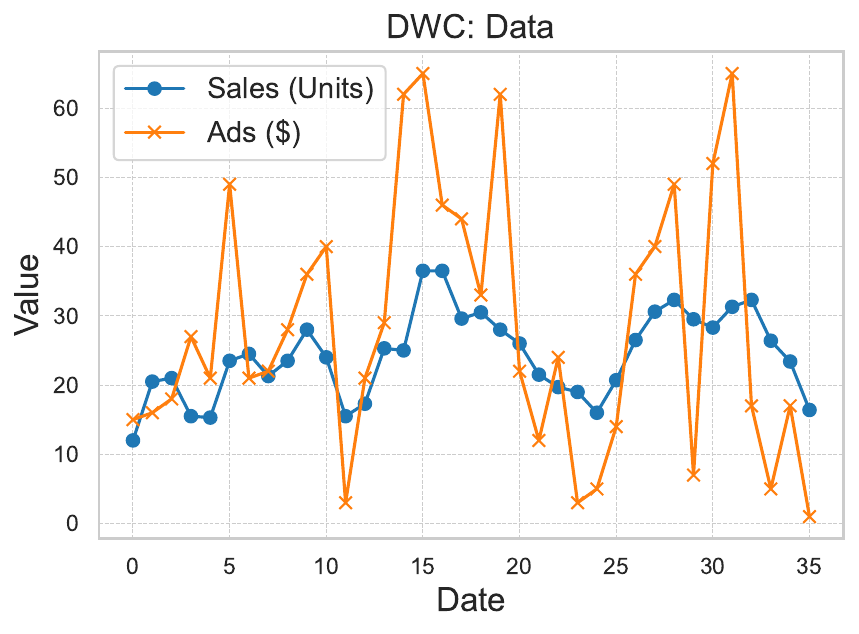} \includegraphics[width=0.4\textwidth]{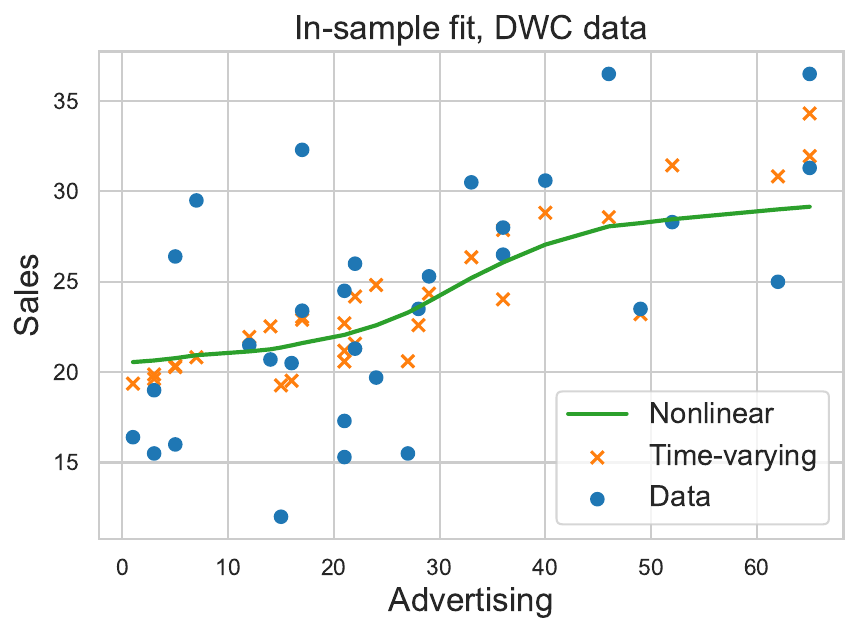}
    \tcaption{DWC, Data and Fit}{At left, the raw DWC sales (in units) and advertising (in dollars) data. At right, the in-sample fit of the nonlinear GP model and the time-varying GP model.}
    \label{fig:dwc_data}
\end{figure}

The DWC data are plotted in the left panel of \Cref{fig:dwc_data}. We analyzed this data assuming no advertising carryover, reflecting \cite{bass1972testing}'s description of the data as ``an excellent candidate for a study of the short-run effect of advertising on sales.''\footnote{This assumption also: (1) performs better in holdout MSE, relative to carryover specifications; and (2) limits the number of initial time periods dropped for computing a stock variable, which is important for such a short time series.} The fits of the time-varying and nonlinear (GP) models are shown in the right panel of \Cref{fig:dwc_data}, and the holdout performances of all four models are summarized in the first two rows of \Cref{tab:dwc_lp_rmse}. From both of these, we see that the models are conflated: from \Cref{fig:dwc_data}, the fits of the two basic GP models are clearly indistinguishable, and from \Cref{tab:dwc_lp_rmse}, the posterior intervals of RMSE for all models are overlapping, suggesting that based on standard analysis, it is not possible to tell the two models apart.

\begin{figure}
    \centering
    \includegraphics[width=0.5\textwidth]{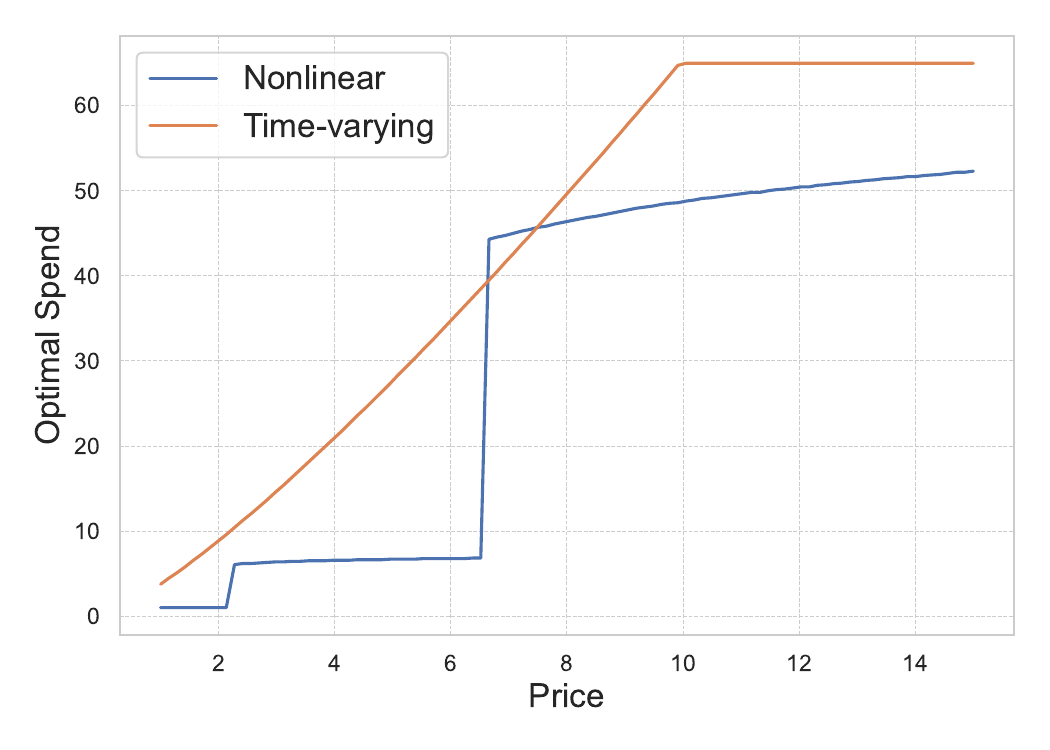}
    \tcaption{DWC, Optimal Spend}{Optimal one-ahead advertising spend levels recommended by the nonlinear and time-varying GP models, as a function of product price.}
    \label{fig:dwc_optimal}
\end{figure}

The more important question is, to what extent does this conflation affect decision-making? For this, we focus just on our two focal GP models (nonlinear and time-varying), with logged inputs, just as in the simulated example previously. To simplify the analysis, we assume for the time-varying GP that the next period elasticity will be the same as the last observed period. One feature of the DWC data is that we observe sales in units, but do not observe prices. This data limitation is also an opportunity: instead of thinking of the optimization problem given a single price point, we conduct the optimization exercise over a grid of hypothetical prices, and see how the different models give different optimal spending recommendations at different revenue levels. That is, for each hypothetical price, we can compute what would be the optimal advertising level implied by each model. The results of this analysis are shown in \Cref{fig:dwc_optimal}, where the hypothetical price level is on the x-axis, and the optimal spending level is on the y-axis. We see that, for the majority of the price range, the time-varying model recommends setting a higher level of spending compared to the nonlinear, sometimes dramatically so. 

\paragraph{Lydia Pinkham (LP)}

\begin{figure}
    \centering
    \includegraphics[width=0.45\textwidth]{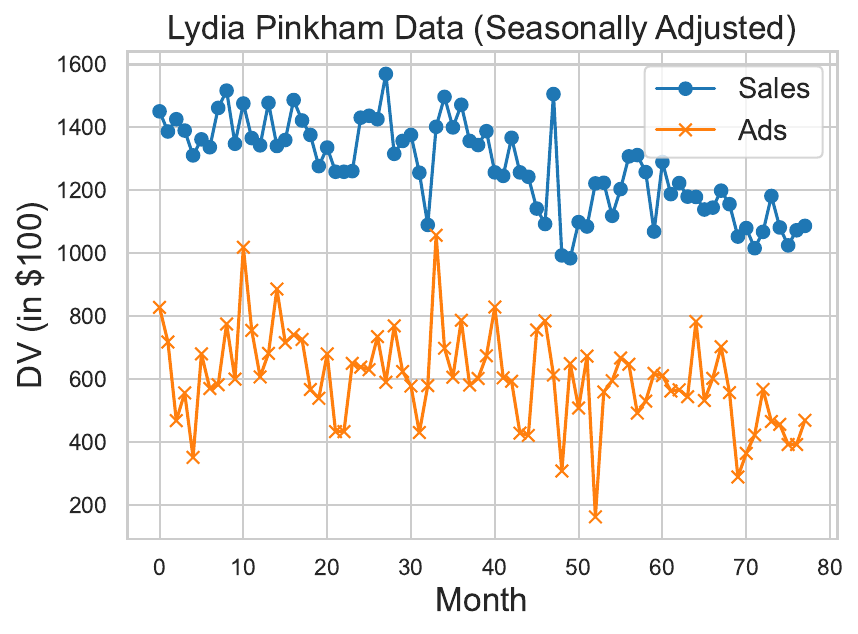} \includegraphics[width=0.45\textwidth]{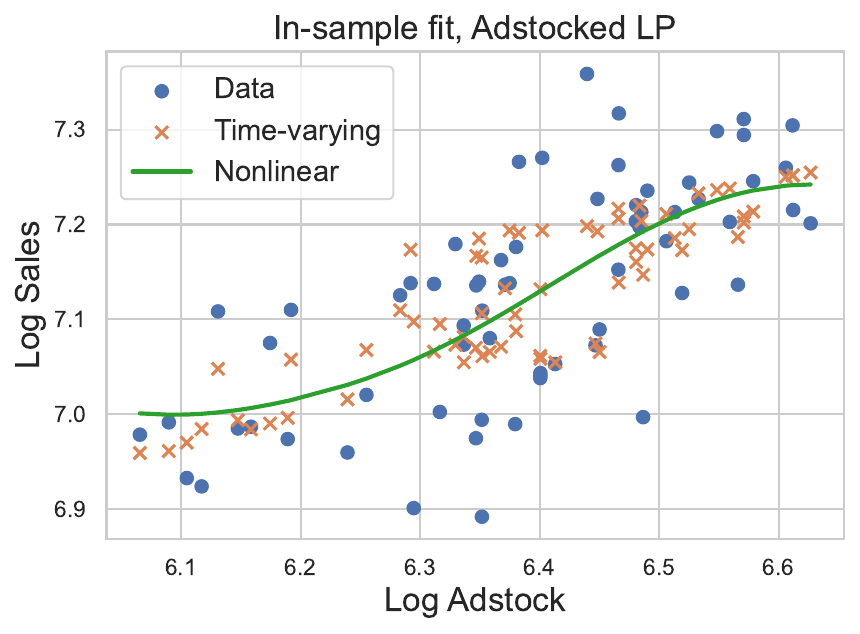}
    \tcaption{LP, Data and Fit}{At left, the raw Lydia Pinkham data (seasonally adjusted), from \cite{palda1965measurement}. At right, the data after applying an AdStock transformation with $L=13$ and $\lambda=0.7$, as suggested in \cite{bultez1979does}. The right panel is overlaid with the fits of the nonlinear (GP) and time-varying models.}
    \label{fig:lps_data_fit}
\end{figure}

The Lydia Pinkham data are plotted in the left panel of \Cref{fig:lps_data_fit}. Specifically, we use the seasonally adjusted version of the data, provided in \cite{palda1965measurement}, which has been frequently used throughout the literature to measure long-run ad effects \citep[e.g.,][]{bultez1979does}. Given the importance of long-run effects, using a stock variable for advertising is important. Thus, to analyze LP, we apply the stock transformation to advertising, as described in \Cref{eq:adstock}, using $L=13$ and $\lambda=0.7$. $L=13$ suggests a 12-month carryover period, which effectively degrades the effect to almost zero. $\lambda=0.7$ is the value suggested in \cite{bultez1979does}, who extensively examined carryover in the LP data. The resulting AdStock data is shown, on the log scale, in the right panel of \Cref{fig:lps_data_fit}, overlaid by the fits of two models: the nonlinear and time-varying GP-based models. 

First, we note that, as before, all models are conflated: the holdout RMSEs from the second section of \Cref{tab:dwc_lp_rmse} are all nearly identical, suggesting a standard analysis could not distinguish the models. Moreover, from \Cref{fig:lps_data_fit}, we see the models fit nearly identically. The nonlinear model exhibits an S-shaped response, as captured by the green line in \Cref{fig:lps_data_fit}, while the time-varying log-linear model implies a positive but decreasing elasticity over the duration of the data.

Crucially, as before, these models give different suggestions for optimal spending. To determine optimal spending, we must generalize our optimization procedure from before to take into account carryover effects, as current period expenditures will affect the future through the stock variables. In the literature on optimal ad spending, the optimization problem with carryover is often considered in the context of some long-run, steady state of advertising \citep[e.g.,][]{bultez1979does}. In our context, deriving such a steady state is infeasible, given the nonparametric nature of our response functions. Instead, we turn to current industry practice, described in \cite{jin2017bayesian}, which mirrors the optimization described previous, but modified to account for carryover. This procedure optimizes spending over a test window, $[T_1, T_2]$, such that $T_2 < T - L$. To account for the carryover effect of advertising, profits are predicted both during the test window, and during the post period, during which time spending is held fixed at its observed values. Mathematically, denoting $\x_t = (x_{1t}, \ldots, x_{Jt})$, this corresponds to:
\begin{equation}
    \max_{\substack{\x_{T_1}, \ldots, \x_{T_2}}} \sum_{t \ge T_1} \left[\hat{y}_t(\x_{T_1}, \ldots, \x_{T_2} \,|\, \x_{t \not\in [T_1, T_2]}) - \sum_j x_{jt} \right], \label{eq:spend_optim_with_carryover}
\end{equation}
where $\hat{y}_t(\x_{T_1}, \ldots, \x_{T_2} \,|\, \x_{t \not\in [T_1, T_2]})$ denotes the predicted revenue in period $t$ under the counterfactual levels of spending in the test periods, $\x_{T_1}, \ldots, \x_{T_2}$, holding all other spend levels fixed. Since the test window is restricted to be at least $L$ periods before the end of the data, this ensures that any carryover effects will be fully realized by the end of the post period. 

Returning to our results on the LP data, we again focus just on the nonlinear and time-varying GP models. We consider a single period optimization window, corresponding to ${[T-L-1, T-L]}$ (i.e., the last period for which we can observe the entire carryover effect). To optimize spending, we consider the range of possible advertising values that correspond to stock values within the range of those observed in the data. This restriction means that the optimization results will not be based on extrapolation. Following this procedure, we find, for the nonlinear model, an optimal ad spend of \$90,127.63, and for the time-varying model, an optimal ad spend of \$78,333.87.\footnote{Note that \$78,333.87 is the lowest value considered: as Lydia Pinkham's true ad spending during the post-test period was lower than all prior periods, \$78,333.87 is the lowest value of test period ad spend that led to test and post-test period AdStocks in the range of the prior data. It is likely that the true optimum under the time-varying model is lower than this.}

\subsection{Modern MMM Data} 

To assess whether the conflation issue persists in modern multichannel data, we assembled weekly level sales and advertising data using NielsenIQ Retail Scanner and Nielsen AdIntel data for 2018--2019. We selected a set of four large, national brands, each coming from a different product category: chocolate, pet food, coffee, and beer.

Three important factors in these modern datasets that were not relevant in the classic datasets are: (1) multiple advertising channels; (2) trend and seasonality effects; and (3) promotions that are unobserved in the data, except as large spikes in spending. Factor (1) is already handled within our general framework. To control for (2), we modify our focal models to include a time-varying intercept:
\begin{equation} 
    y_t = \alpha(t) + g_1(x_{1t}, t) + g_2(x_{2t}, t) + \ldots + g_J(x_{Jt}, t) + \varepsilon_{t},
\end{equation}
where again the $g_j$ are set either to be nonlinear and static, or time-varying and (log-)linear. To model $\alpha(t)$, we again use a GP prior, but this time with a slightly more sophisticated ``Trend-Season'' (TS) covariance function. This kernel is the sum of a squared exponential kernel, and a periodic kernel, which captures regularly recurring patterns of a given cycle length $c$ \citep{williams2006gaussian}. As sums of kernels are, themselves, kernels, this is a valid kernel. Mathematically, this corresponds to:
\begin{gather}
    \alpha(t) \sim \mathcal{GP}\left(0, k_\mathrm{TS}(t, t'; \btheta)\right) \\
    k_\mathrm{TS}(t,t'; \btheta) = \eta_\mathrm{T}^2 \exp\left(-\frac{(t-t')^2}{2\rho_\mathrm{T}^2}\right) + 
    \eta_\mathrm{S}^2 \exp\left(-2\frac{\sin^2(\pi |t-t'| / c)}{\rho_\mathrm{S}^2}\right).
\end{gather}
This specification allows for us to parsimoniously capture both trend effects, of a smoothness determined by $\rho_\mathrm{T}$, and seasonality effects of cycle $c$. In addition, we include dummies for major US holidays. To capture (3), we construct a ``promotions'' dummy variable that captures sparse, sporadic spending on certain channels (i.e., channels wherein spending happens only rarely, in single bursts), and unusual spikes of spending in the main advertising channels.

With these modifications, our analysis proceeds as before: for each dataset, we fit competing models, leaving some data heldout. For this analysis, we focus just on the nonlinear and time-varying GP models. Prior to estimating the models, we apply the AdStock transformation, setting the decay parameter $\alpha = 0.3$ \citep[consistent with industry practice, e.g.,][]{robyn} and the number of lags $L=13$ as suggested in \cite{jin2017bayesian}. Finally, we log the adstocked data. We then predict the heldout data, and compute the posterior RMSE, checking for model conflation. Finally, we explore the managerial implications of the conflation, if it exists. Rather than presenting the results from each analyzed dataset, we instead present the combined RMSE conflation results in \Cref{tab:nielsen_rmse}, and then focus on a single case study below: a brand in the coffee category.

\begin{table}
\centering
\tcaption{Nielsen Data --- RMSE Comparison}{Posterior mean RMSEs for each dataset from Nielsen examples, together with 95\% credible intervals in brackets.}
\label{tab:nielsen_rmse}
\resizebox{0.9\textwidth}{!}{%
\begin{tabular}{lcccc}
\toprule
 & \text{Chocolate} & \text{Pet Food} & \text{Coffee} & \text{Beer} \\
\midrule
\text{Nonlinear (GP)} & 368,766 & 48,438 & 722,076 & 443,817 \\
 & [207,634, 601,163] & [27,920, 83,249] & [509,527, 919,820] & [236,730, 911,732] \\[1ex]
\text{Time-varying (GP)} & 387,623 & 41,825 & 733,269 & 423,085 \\
 & [207,350, 671,246] & [24,631, 65,232] & [490,785, 994,303] & [207,914, 802,740] \\
\bottomrule
\end{tabular}}
\end{table}

\begin{figure}
    \centering
    \includegraphics[width=0.7\textwidth]{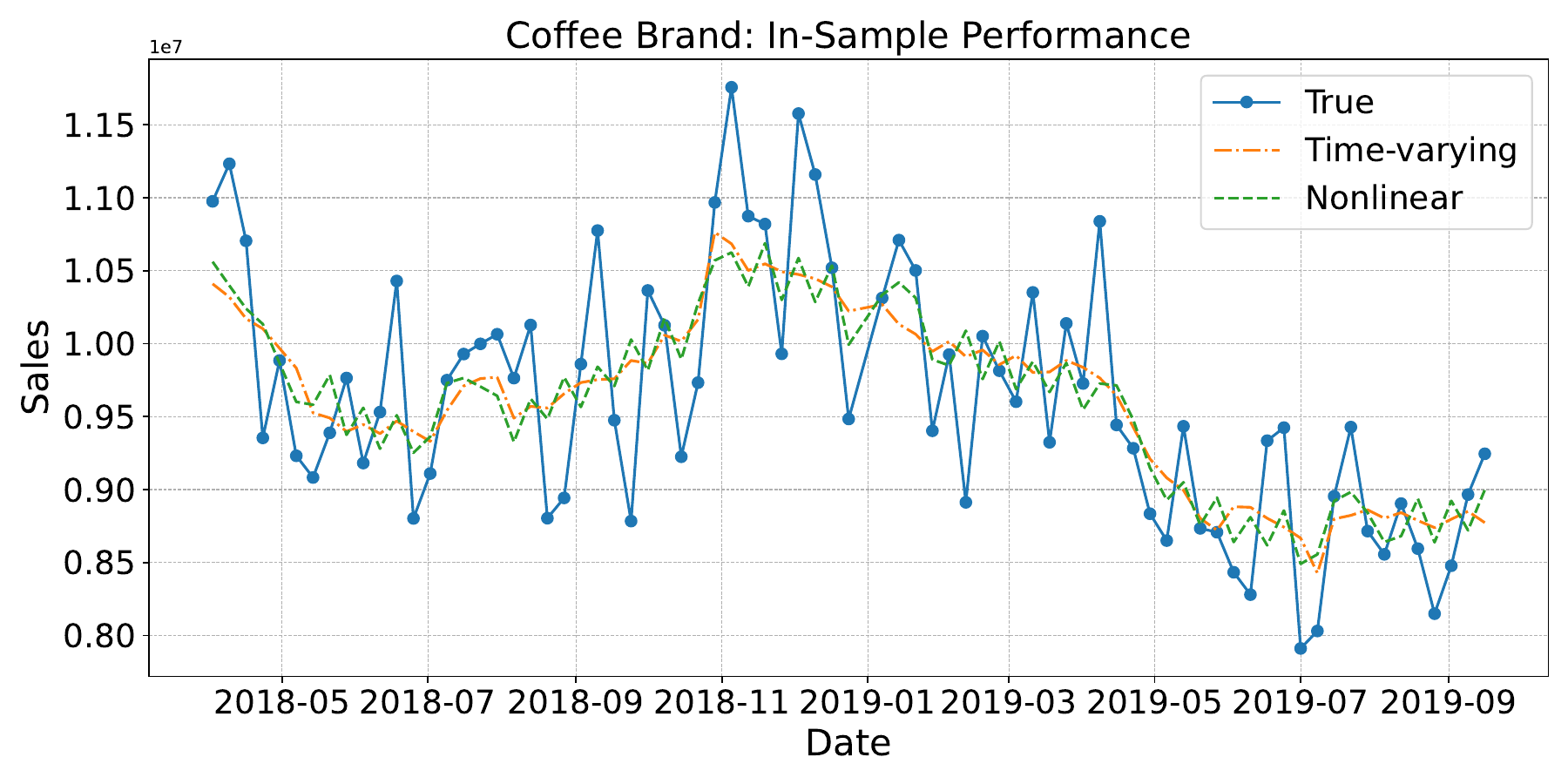}
    \tcaption{Coffee brand, fit}{In-sample fit of nonlinear GP and time-varying GP models (all include trend and seasonality, promotions, and holidays).}
    \label{fig:coffee_fit}
\end{figure}

\begin{figure}
    \centering
    \includegraphics[width=0.4\textwidth]{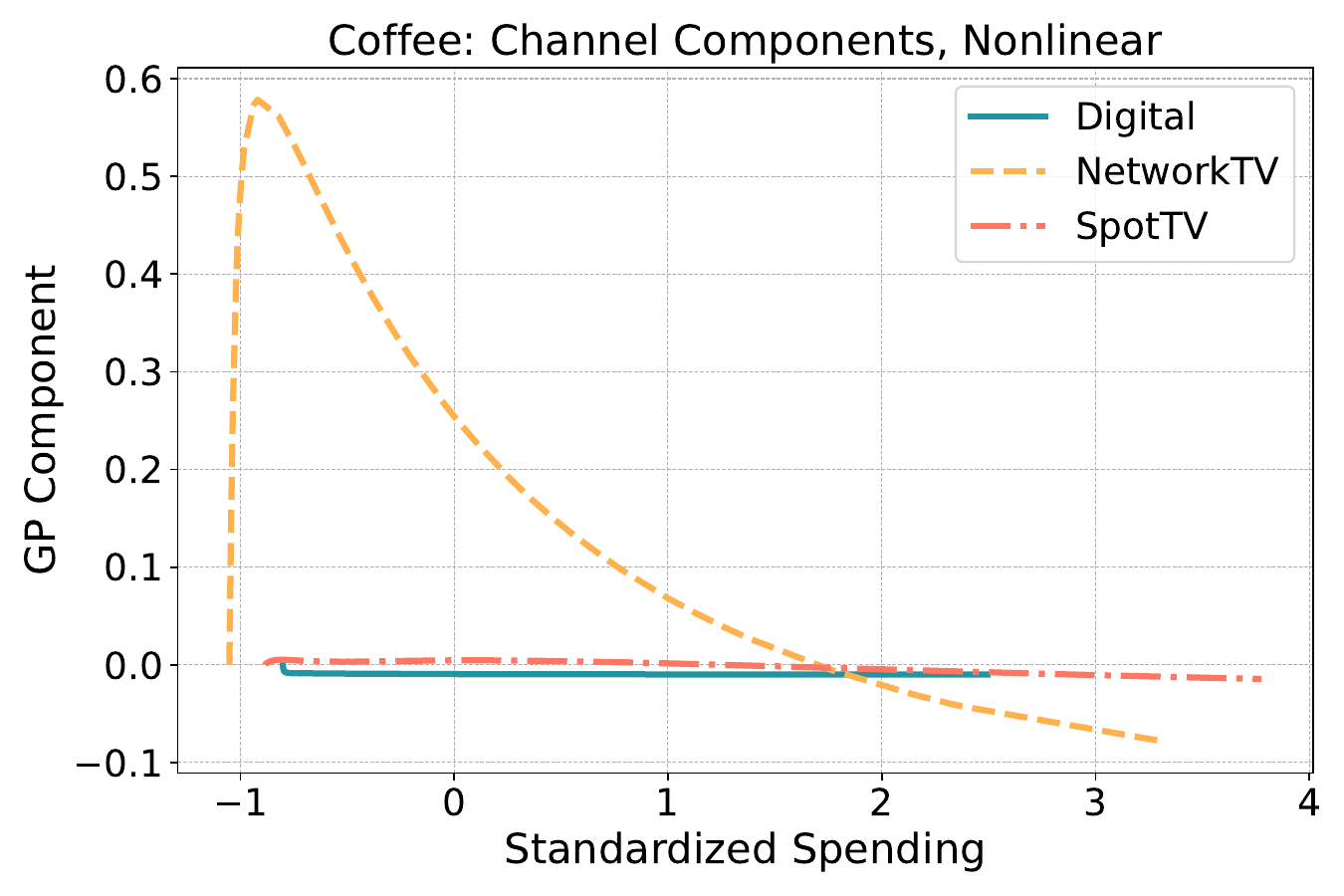} \includegraphics[width=0.4\textwidth]{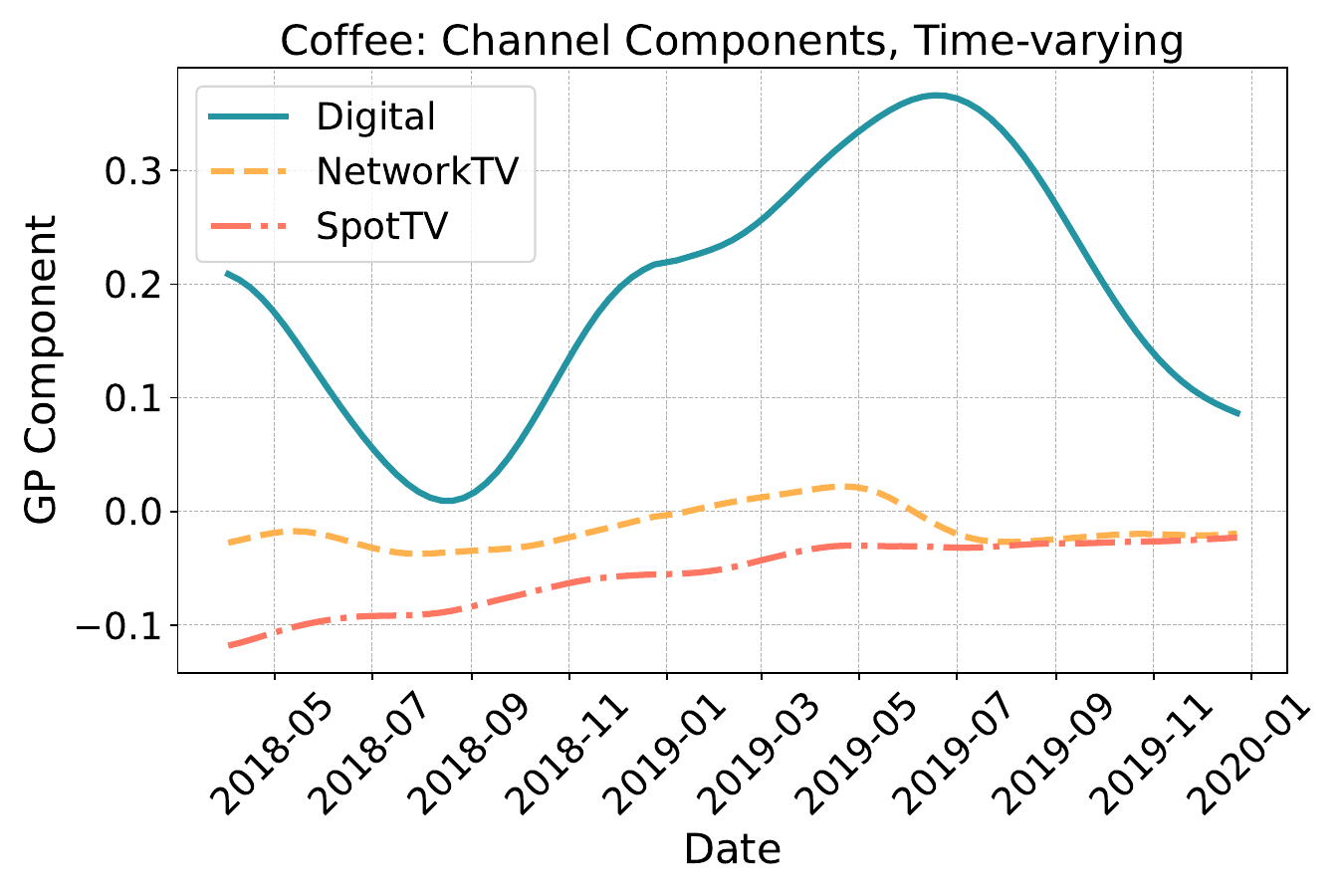}
    \tcaption{Coffee brand, GP components}{At left, GP channel components from the nonlinear GP model, over standardized spending. Response curves are normalized to start at zero. At right, GP coefficients by channel from the time-varying GP model, over time.}
    \label{fig:coffee_coefs}
\end{figure}

\paragraph{Results} First and foremost, we can see that, again, there is conflation in all cases: from \Cref{tab:nielsen_rmse}, we see that the posterior intervals are all coinciding, for all datasets, suggesting again that standard model selection practices would not be able to select the correct model. Focusing on the case of our coffee brand in \Cref{fig:coffee_fit,fig:coffee_coefs}, we see that, as before, both models fit extremely well, while having very different estimates of each channels' effectiveness at different points in time. We also note that, in this case (and indeed, in all of the modern cases considered), a huge amount of the variation in revenue is captured by the trend and seasonality component of the model, not by marketing spend. 

\paragraph{Case Study: Spend Optimization} Finally, we again consider what these different models imply about budget allocation. To do so, we utilize the estimates from the nonlinear and time-varying GP models (\Cref{fig:coffee_coefs}), and follow the same approach to spend optimization described previously. Specifically, we focus on redistributing the actual advertising budget in the last week of September 2019 to allow for carryover effects. We construct candidate spending allocations by varying the three main channels as a percentage of the total actual spend in increments of 5\% (e.g., 20\% Digital, 55\% Network TV, 25\% Spot TV). We then remove the candidate allocations that assign to at least one of the channels a level of spending that falls outside of the historical range.\footnote{This ensures that our optimal solution is not relying purely on extrapolation of the response function.} For the remaining allocations, we compute the implied AdStocks, treating the actual spending in the periods before and after the target period as given. We predict the sales in the target period and the following 13 weeks (our chosen number of lags for the AdStock) to account for carryover effects using the nonlinear GP and time-varying GP models, and choose the allocation that yields the highest total revenue over the 14 periods (the total spending is by construction identical for all candidate allocations). The results are presented in \Cref{tab:coffee_optim}.

\begin{table}
    \centering
    \begin{tabular}{lccc|c}
    \toprule
         & Digital & Network TV & Spot TV & Conflation Cost \\ \midrule
     Nonlinear GP    & 15\% & 50\% & 35\% & \$227,000 \\
     Time-varying GP    & 15\% & 85\% & 0\% & \$61,000 \\ \bottomrule
    \end{tabular}
    \tcaption{Coffee: Optimal Allocation}{Optimal allocation of the total budget in the last week of September 2019 across three main channels. The last column presents the cost of conflation: the predicted difference in revenue between the nonlinear GP and the time-varying GP optimal allocations, given that one of them is the true DGP.}
    \label{tab:coffee_optim}
\end{table}

We can see that the two models agree on the optimal share of the Digital channel (15\%). Note that 15\% of spending on digital was the maximum observed in the data, which was a constraint we imposed. However, they drastically disagree on the composition of Network and Spot TV: while the nonlinear GP model advises to allocate $35\%$ of the budget to Spot TV, the time-varying GP model suggests not to spend on Spot TV at all. These allocations stem directly from the estimates in \Cref{fig:coffee_coefs}: we see that, for nonlinear, there is an estimated peak in the effect of Network TV, after which the model estimates a decreasing effect.\footnote{While such an effect may be a priori surprising, negative and non-increasing effects are, in fact, commonly observed in marketing effectiveness estimates based off this same data. \cite{shapiro2021tv}, for instance, find a number of negative effects in their study of advertising. Moreover, a non-monotone effect like ours may result in a negative overall effect when the model does not allow for a flexible functional relationship, as our model does.} Thus, the nonlinear model suggests spending some, but not too much, on Network TV. On the other hand, for the time-varying model, the only effect is that, for almost all time periods, Network TV's elasticity is higher than Spot TV's, leading to the observed allocation. The last column of \Cref{tab:coffee_optim} quantifies the cost of conflation: if the nonlinear GP is the true model, then choosing the optimal allocation assigned by the time-varying GP model in the last week of September 2019 will lead to a \$227,000 loss over the next 14 weeks, relative to using the nonlinear model. Vice versa, if the time-varying GP is the true model, then choosing the optimal allocation assigned by the nonlinear GP model will result in a \$61,000 loss. In other words, model misspecification can have sizable economic consequences even under fairly strong restrictions: we varied only channel allocation, not the size of the spending, and in only one period.

\section{Separation Tests}

We have shown that nonlinear and time-varying effects in marketing mix models are often conflated, potentially resulting in diverging recommendations on how much a company should spend on advertising. This naturally raises the issue of whether anything can be done to separate the two effects and identify the more appropriate model for optimizing spending. In this section, we introduce a way to test which of the two types of effects better reflects the true underlying response function. We do so by developing what we term \textit{separation tests}, which are spending policies designed to break the conflation mechanisms discussed previously, and allow simple holdout metrics to determine which of the models is correct. Specifically, we introduce two tests: we start with a decision-theoretic approach that designs a test wherein spending levels are set to maximize the expected difference in the predictions between the nonlinear and time-varying models. While this approach is most powerful, in terms of having a high probability of truly separating the effects, it is also hard to implement. Thus, we also provide an easy-to-implement ``seesaw'' test that can, under some conditions, achieve separation without such complexities.

\subsection{Best One-shot Test: Maximal Separation}

The intuition behind our first proposed intervention on spending is straightforward: we select the level of spending in a single period that maximizes our expected ability to tell the two models apart. To do this, we consider a range of possible spending values, predict the sales in the next period using the contender models, as if we were planning to set the advertising spend, and then set spending in the next period to the level with the highest difference between the two predictions. Then, following the test, we assess which prediction was actually closest to the truth.

More formally, assume there exist historical data, $D_T=\lbrace x_t,y_t\rbrace^T_{t=1} $, on which both models have been estimated (i.e., the nonlinear model from \Cref{eq:mv_nl_model} and the time-varying model from \Cref{eq:mv_dyn_model}). For ease of exposition, throughout this section, we assume there is only a single spending channel, though the ideas generalize to multiple channels. To separate the two models, we first consider a set of possible spending levels, $\lbrace x^p_1,\ldots,x^p_N\rbrace$. Then, for each of the period of the test $(t\geq T)$, we set $x_{t+1}$ to the candidate level that maximizes the difference between the predictions from the two models, among all candidate spending levels:
\begin{equation}
    x_{t+1} = \underset{x\in\lbrace x^p_1,\ldots,x^p_N\rbrace}{\arg\max} \left\vert \hat{y}^{NL}_{t+1}(x|D_t) - \hat{y}^{TV}_{t+1}(x|D_t)\right\vert. \label{eq:max_disentangling_rule}    
\end{equation}
Here, $\hat{y}^{NL}_{t+1}(x|D_t)=E_{f}(f(x)|D_t)$ is the expected posterior predicted revenue from spending $x$ under the nonlinear model, and $\hat{y}^{TV}_{t+1}(x|D_t)=E_{\beta_{t+1}}(\beta_{t+1}\cdot  x|D_t)$ is the expected posterior predicted revenue for period $t+1$ from spending $x$ under the time-varying model. We term the maximum from \Cref{eq:max_disentangling_rule} the \textit{maximal separation} between the two models, and thus this test, the \textit{maximal separation test}.

\begin{figure}
    \centering
    \includegraphics[width=0.4\textwidth]{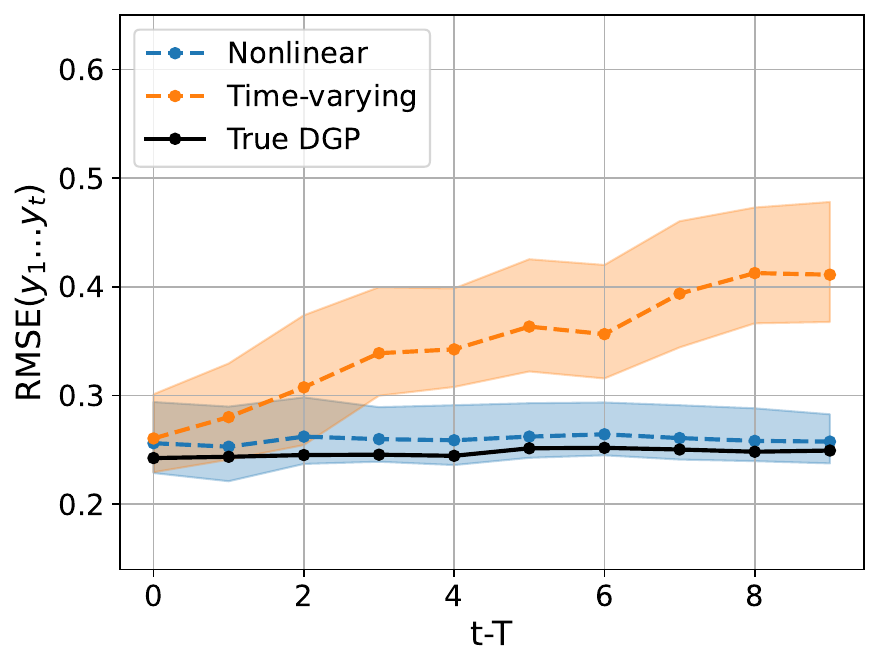} \includegraphics[width=0.4\textwidth]{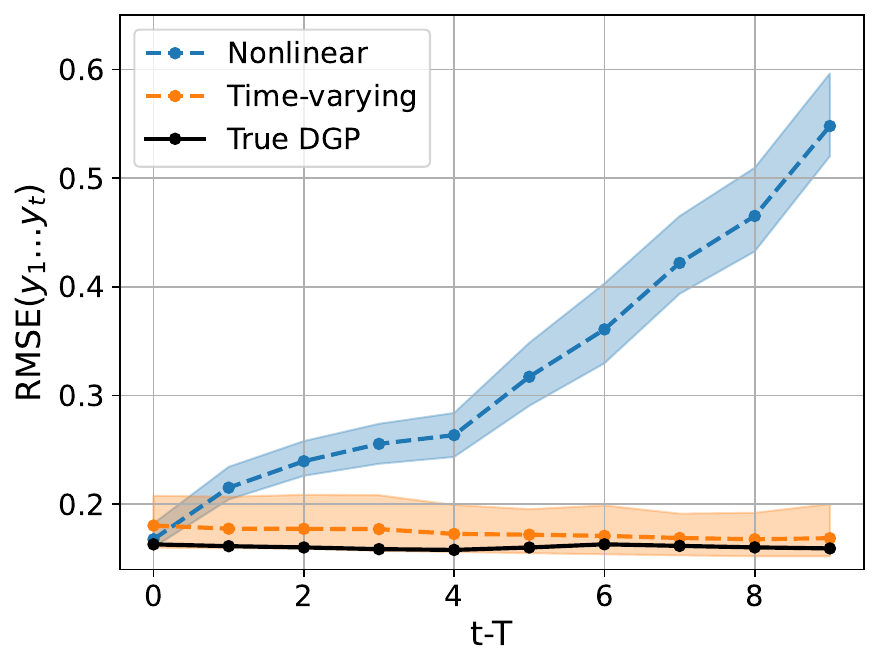} 
    \tcaption{In-sample RMSE as a function of maximal separation periods}{At left, the true DGP is nonlinear. At right, the true DGP is time-varying.}
    \label{fig:maximal_in_sample}
\end{figure}

To illustrate, we apply the test to two simulated examples: (1) a time-varying DGP, using the same dynamic effects as in our illustrative example from the introduction, and (2) a nonlinear DGP, where the nonlinearity is a simple log function. In both cases, we generate 48 periods of data (simulating four years of monthly data), then fit both models to both datasets. Then, starting in period 49, for each DGP, we use the maximal separation test to set spending, as in \Cref{eq:max_disentangling_rule}. \Cref{fig:maximal_in_sample} shows how the in-sample RMSE changes for each DGP as a function of the number of periods of the test. At test period 0, meaning before the test begins, we see that, in both cases, the models are conflated (i.e., the posterior RMSE distributions are overlapping). However, after two periods of maximal separation testing, the posterior intervals of the two models' RMSEs no longer overlap, with the true model always having the lower RMSE. Note that this separation happens (marginally) faster when the true DGP is time-varying, which is consistent with our theory, as the conditions for the nonlinear model to approximate the time-varying are stronger. Crucially, we note that, in both cases, separation happens quickly, suggesting that a firm need only conduct this test for 1-2 periods to enable some degree of separation.

\begin{figure}[!h]
    \centering
    \includegraphics[width=0.8\textwidth]{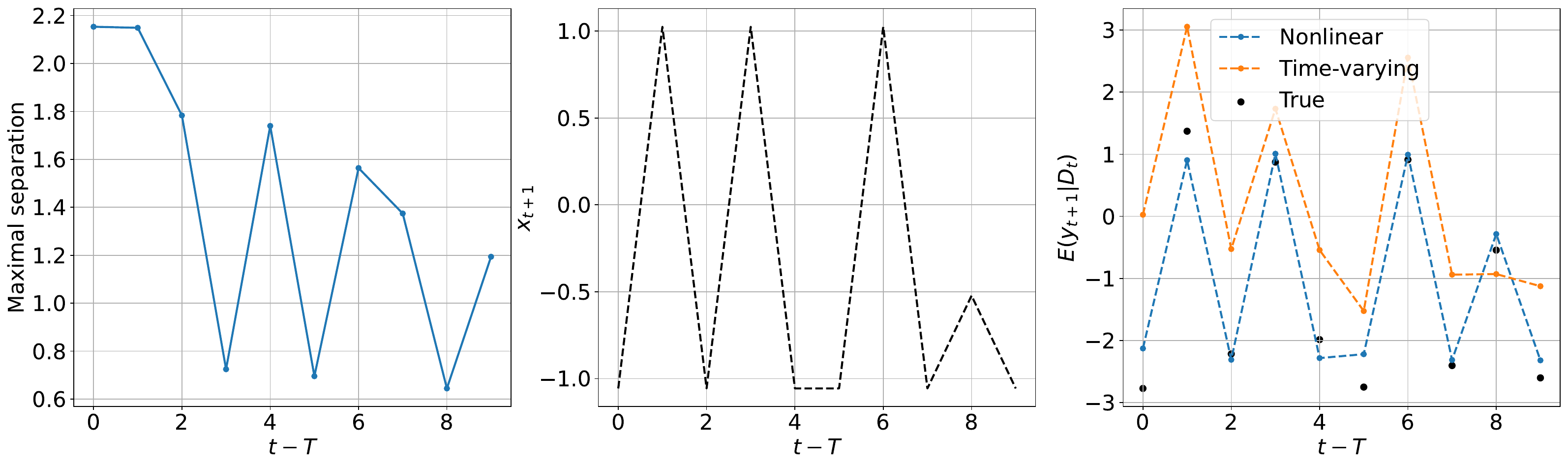} 
    \includegraphics[width=0.8\textwidth]{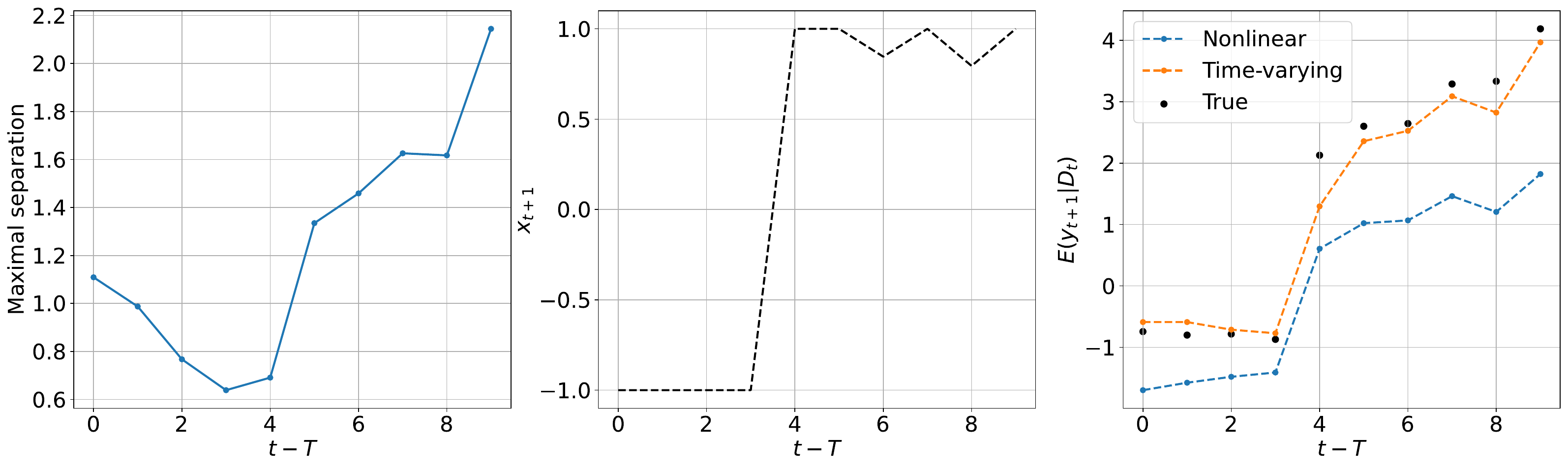} 
    \tcaption{Evolution of maximal separation procedure}{In the top row, the true DGP is nonlinear, while in the bottom row, the true DGP is time-varying. The left panel shows the maximal separation between the models at each step of applying the maximal separation test. The middle panel shows the optimal spending level at each step. The right panel shows the predictions made by each model evaluated at the optimal spending level $x_{t+1}$. }
    \label{fig:maximal_evolution}
\end{figure}

To better understand how the test works, in \Cref{fig:maximal_evolution}, we plot, for each period during the test: (1) the maximal level of separation between the two models; (2) the spending level $x_t$ that achieves maximal separation; (3) the two models' predictions for that value of $x_t$. The dynamics in how the test unfolds shed light on what is happening: over time, the maximally separating value of $x_t$ changes, as the incorrect specification adjusts to this new pattern of spending, and learns to fit the response better. The maximal separation test responds by finding a new value to test the models. This adversarial nature of the algorithm allows it to quickly and efficiently identify the correct specification. 

In the top panel in \Cref{fig:maximal_evolution}, where the true DGP is nonlinear, we see the maximally separating pattern of spending involves alternating between high and low values. This pattern makes sense: the regularization built in to the time-varying model implies that, within a short time period, the response to spending should follow (roughly) a linear model. The truth, on the other hand, is nonlinear. Thus, under a short time frame, the approximately linear model implied by the time-varying model cannot capture the oscillating response to the two extreme spending levels. This is consistent with our theory: smooth dynamics can flexibly capture a nonlinear static DGP when the curvature of the nonlinear function and the local variability of spending are both low (\Cref{eq:ols_regularizer}). The seesaw spending pattern causes high local variability in spending, in turn causing the local linear approximation on which the time-varying model relies to become less accurate, separating the two models.

Interestingly, the same process evolves differently when the underlying data-generating process is time-varying but linear (bottom panel in \Cref{fig:maximal_evolution}). The maximally separating level of spending stays stable for some time in order to exploit underlying response dynamics that the time-varying model can easily accommodate, while the nonlinear cannot. Intuitively, under the time-varying model, the same level of spending can lead to different responses, as the elasticity evolves. The same is not possible under a nonlinear model. Thus, by merely keeping spending constant for some time at the initial maximally separating value, the test quickly breaks the conflation between the two models. This is also consistent with our theory: the conditions under which the nonlinear model approximates the time-varying are more strict, and thus, separation is easier to achieve.

\subsection{Separation Heuristic: Seesaw Test}

The policy defined in \Cref{eq:max_disentangling_rule} may not always be easy to implement, since it requires evaluating both models' predictions at multiple hypothetical spending levels. Thus, we also propose a simpler spending strategy that approximates our decision-theoretic approach, and is rooted in breaking the conditions for conflation outlined in our theory. The analysis of the dynamics in the maximal separation test highlight that separation is more difficult when the true model is nonlinear. Thus, for designing an easy, generic separation test, we focus on this harder case. 

Specifically, inspired by the dynamics in \Cref{fig:maximal_evolution}, we propose a test that involves alternating spending between high and low spending levels, which we term a \textit{seesaw test}. Mathematically, the seesaw test involves alternating spending between $x^H=\max\lbrace x^p_1,\ldots,x^p_N\rbrace $ and $x^L=\min\lbrace x^p_1,\ldots,x^p_N\rbrace$, one period at a time: $x_{T+1} = x^H, x_{T+2} = x^L, x_{T+3} = x^H, x_{T+4} = x^L$, and so on. Such a test can be thought of as a type of incrementality test, similar to what is referred to in practice as a ``bump up'' test.\footnote{Such a test also resembles a common advertising strategy --- pulsing --- which has been shown to be both common and, in some cases, optimal, in a number of past marketing studies \citep[e.g.,][]{feinberg1992pulsing}.} These tests are often used to assess the marginal effectiveness of a particular channel. Such a strategy combines features that break the approximation mechanisms for both the nonlinear and time-varying models: first, by alternating spending between high and low levels, $x_t$ is returning to similar levels over and over. If the true relationship is static and nonlinear, sales should also go back to the same level repeatedly (after accounting for seasonality). On the other hand, if the seesaw test is done within a short period of time, then under a time-varying model, the elasticity implied by such a test should remain roughly constant, allowing us to observe the suitability of the time-varying model. 

\begin{figure}
    \centering
    \includegraphics[width=0.4\textwidth]{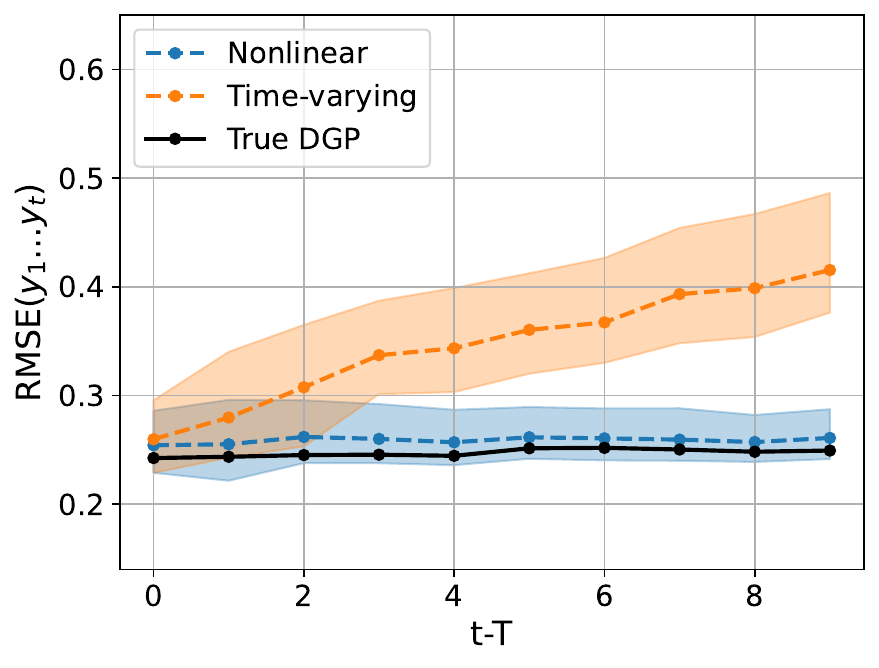} \includegraphics[width=0.4\textwidth]{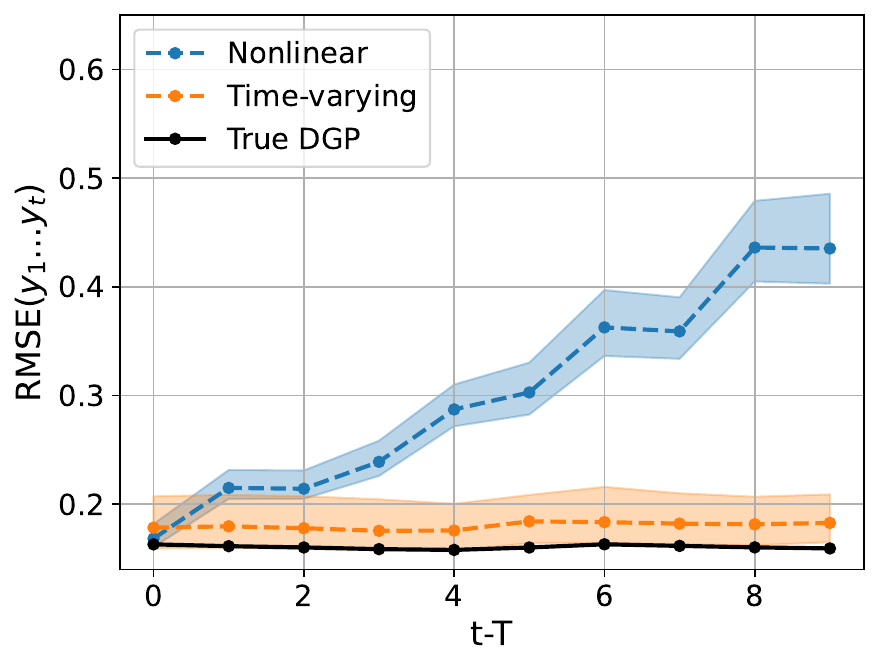} 
    \tcaption{In sample RMSE as a function of seesaw periods}{At left, the true DGP is nonlinear. At right, the true DGP is time-varying.}
    \label{fig:seesaw_in_sample}
\end{figure}

To evaluate the usefulness of the seesaw test, we again turn to the two examples described previously: the time-varying DGP from the introduction, and a simple nonlinear DGP with a log nonlinearity. We applied the seesaw test in a synthetic spending simulation using these DGPs, just as we did with the maximal separation test, fitting both models to both DGPs using 48 periods of data, then implementing the test going forward. We plot the resulting two RMSE patterns during the testing period in  \Cref{fig:seesaw_in_sample}. As the figure shows, the RMSEs of the two models diverge after a few intervention periods, highlighting the ability of the seesaw test to approximate the maximal separation test, and identify the correct model. While these results suggest the viability of the seesaw test, it is worth noting, that the seesaw test may not always be as effective as the maximal separation test, particularly in contexts with strong carryover effects and long lags. In these cases, even if spending follows a seesaw pattern, the smoothing induced by stocking variables may limit how much variability can be induced in a short period of time. In these cases, a direct implementation of the maximal separation test may be preferable. 

\section{Conclusion}

In this paper, we documented an important issue of model identification in modern marketing mix models. As practitioners attempt to capture increasingly complex effects in MMMs, like nonlinearities and dynamics, our results suggest caution is warranted: the simple data used for building such models often cannot uniquely identify such complexity. We demonstrated these issues through theory, simulation, and applications to marketing mix datasets, both classic and contemporary. In doing so, we also introduced a family of Bayesian nonparametric marketing mix models that can flexibly model nonlinear and time-varying effects, whenever such complexity is warranted. Finally, we proposed a series of tests that are straightforward to implement by firms, that can separate out time-varying and nonlinear effects. By implementing such tests in standard MMM practice, our results suggest the risk of conflating effects and making suboptimal decisions diminishes. 

While our work provides what we believe is an important framework for thinking about nonlinearities and time-varying effects, there are also many ways in which our framework could be extended and improved. For instance, including informative mean functions in the GPs (e.g., Hill), encoding monotonicity contraints in the ad response functions, and more thorough use of prior information could all help induce further regularization on the learned functions, and better pin down marketing mix effects. While the minimal assumptions used herein is, we believe, most appropriate for studying the issue of conflation, in practice, our proposed framework might be usefully extended. 

In focusing on the conflation between nonlinear and time-varying effects, our work presently leaves several areas unexplored. First, in real data, there are likely both types of effects at work. In such cases, our work is still relevant, and can be seen as a careful examination of how each type of effect (or the dominant type of effect) might be identified. The GP-based framework we introduced could hypothetically handle such an extension, though standard MMM data may be insufficient for estimating such a complex model. Second, we have not yet explored the issue of how endogeneity intersects with all of these issues. The issues examined here exist even with purely exogenous spending data, and would likely be compounded further with endogeneity issues and corrections. Finally, in industry, there are many common practices that relate to our work, which we leave unexplored. For instance, rather than estimating truly time-varying models, common practice is to estimate MMM models on rolling windows, giving them periodic ``refreshes.'' Such efforts are effectively estimating time-varying effects, and we believe, still result in conflation. In fact, we posit that changing effectiveness observed under refreshes may be an artifact of exploring different areas of a nonlinear response at different points in time. We leave further exploration of this issue to our ongoing research. 

Our work has several implications for managerial practice. Most importantly, we show that managers should not blindly trust existing MMM solutions, and must take care in interpreting their results, as the aggregated data used for MMM estimation often cannot support identification of complex effects. We also encourage managers to incorporate experimentation in their budget setting practices. Both Meta's \citep{robyn} and Google's \citep{meridian} most advanced marketing mix models directly recommend incorporating incrementality tests in advertising spending, to mitigate endogeneity issues and calibrate MMM parameters. We provide another reason for why experiments are important: a carefully designed experiment can also help disentangle nonlinear and time-varying effects and choose the right model specification.

\clearpage
\singlespacing
\bibliographystyle{apalike}
\bibliography{mmm}

\clearpage
\begin{appendices}
\onehalfspacing

\end{appendices}

\end{document}